\documentclass[journal=jctcce,manuscript=article]{achemso}

\usepackage[version=3]{mhchem} 
\usepackage[T1]{fontenc}       
\usepackage{color}
\usepackage{subcaption}
\usepackage{gensymb}
\usepackage{xr}

\newcommand*\sref[1]{%
    S\ref{#1}}

\author{Joaquim Jornet-Somoza}
\affiliation[University of the Basque Country]
{Nano-Bio Spectroscopy Group and ETSF Scientific Development Centre, Department of Materials Physics, University of the Basque Country, CFM CSIC-UPV/EHU-MPC and DIPC, Tolosa Hiribidea 72, E-20018 Donostia-San Sebasti\'an}
\alsoaffiliation[Max Planck Institute for the Structure and Dynamics of Matter]
{Theory Department, Max Planck Institute for the Structure and Dynamics of Matter and Center for Free-Electron Laser Science, Luruper Chaussee 149, 22761 Hamburg, Germany
}
\email{j.jornet.somoza@gmail.com}

\author{Irina Lebedeva}
\affiliation[University of the Basque Country]
{Nano-Bio Spectroscopy Group and ETSF Scientific Development Centre, Department of Materials Physics, University of the Basque Country, CFM CSIC-UPV/EHU-MPC and DIPC, Tolosa Hiribidea 72, E-20018 Donostia-San Sebasti\'an}

\title[MOSTOPHOS-Method]
{Real-Time Propagation TDDFT and Density Analysis for Exciton Coupling Calculations in Large Systems}

\abbreviations{IR,NMR,UV}

\begin{document}

\begin{tocentry}
\includegraphics[width=1.\textwidth]{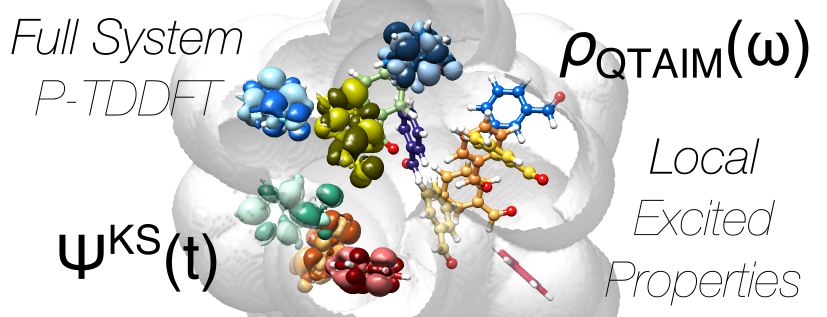}





\end{tocentry}

\begin{abstract}
  {Photo-active systems are characterized by their capacity of absorbing the energy of light and transforming it. Usually, more than one chromophore is involved in the light absorption and excitation transport processes in complex systems.
  Linear-Response Time-Dependent Density Functional (LR-TDDFT) is commonly used to identify excitation energies and transition properties by solving well-known Casida's equation for single molecules. However, in practice LR-TDDFT  presents some disadvantages when dealing with multichromophore systems due to the increasing size of the electron-hole pairwise basis required for accurate evaluation of the absorption spectrum. 
  In this work, we extend our local density decomposition method that enables to disentangle individual contributions into the absorption spectrum to computation of exciton dynamic properties, such as exciton coupling parameters.
  We derive an analytical expression for the transition density from Real-Time Propagation TDDFT (P-TDDFT) based on Linear Response theorems. 
  We demonstrate the validity of our method to determine transition dipole moments, transition densities and exciton coupling for systems of increasing complexity. 
  We start from the isolated benzaldehyde molecule, perform a distance analysis for $\pi$-stacked dimers and finally  map the exciton coupling for a 14 benzaldehyde cluster.}

\end{abstract}

\section{Introduction}

In the last decades the interest to use the natural sun light for an energy transition towards green and clean energy sources has increased. Researchers have focused their investigations on the design of new devices to harvest and use this absorbed light.\cite{Kundu2017,Guo2018} The challenge is to create new light harvesting complexes that can transfer the light energy to a desired reaction center,\cite{Levi2015} such as  natural light harvesting complexes (LHC) \cite{Croce2014,Lambert2012,Scholes2011,Rozzi2013}, or to design new emitters \cite{Adachi2000, Adachi2001} for light and flexible organic light emitting diodes (OLEDs) \cite{Forrest2004,Hung2002,Gustafsson1992} with the quantum efficiency close to unity. Such a quantum efficiency means that all the light energy is transferred to the reaction center for the former, or that all input energy is emitted in the form of light  without dissipation in OLEDs.

These molecular systems usually contain a large number of chromophores, i.e. molecules that can absorb a photon in the UV-Visible region. Their capacity of absorbing and transferring the corresponding energy are the key factors that determine the efficiency of the system. 

From the theoretical point of view, characterization of excited states can be done addressing the eigenvalue problem  either by solving the Schr\"odinger equation using a multiconfigurational ansatz\cite{Szabo1967} or the single-particle Kohn-Sham equations in the time-dependent formalism of the density functional theory (TDDFT) .\cite{Marques2012,Ullrich2012}  An excited state is characterized by its transition dipole moment and the corresponding transition density. The former gives the information about the probability of exciting an electron from the ground state and the most efficient polarization direction of the light, while the transition density informs about the change of the electronic density from the ground state to the excited state. Both of these properties are extremely sensitive to the polarization induced by the environment, and the spectroscopic fingerprint of the isolated molecule in vacuum is usually not sufficient even for a qualitative description.\cite{Mennucci2011, Morzan2018}

The common way to obtain the transition dipole moment for a given excitation in the TDDFT framework is to solve the linear-response Casida's equation.\cite{Casida1995} For that, one needs to compute the ground state Kohn-Sham (KS) wave function. In addition, this procedure requires a large number of well converged unoccupied (or virtual) KS states in order to have a good representation of the electron-hole pair transition space. Nevertheless, this procedure has two main drawbacks: i) large molecules / complex systems require a large set of KS orbitals which makes the solution of the Casida's equation unfeasible, ii) high energy excitations are usually not properly described due to the lack of well converged high energy virtual KS orbitals. 

Several approaches have been developed to reduce the number of electron-hole pair transitions needed to solve the Casida's equation (e.g. subsystems TDDFT,\cite{Neugebauer2007} simplified Tamm-Dancoff density functional approach,\cite{Grimme2013} simplified TDDFT,\cite{Bannwarth2014} tight-binding approaches to TDDFT (TD-DFTB,\cite{Niehaus2001} TD-DFT+TB\cite{Ruger2016}), but they make some approximation on the environment interaction and keep failing for high energy excitations.
On the other hand, Real-Time Propagation TDDFT (P-TDDFT) provides an efficient alternative to these methods. It has been demonstrated that P-TDDFT is an excellent platform for studies of such properties and processes as molecular\cite{Theilhaber1992} and electron\cite{Lopata2011} dynamics, linear and non-linear optics,\cite{Takimoto2007} transport properties,\cite{Evans2009} single and triplet excitations,\cite{Isborn2009, Oliveira2008} dynamical hyperpolarizabilites,\cite{Ding2013} and exciton decay dynamics.\cite{Peng2015} But probably, the most useful advantage that P-TDDFT offers is the possibility to obtain all frequency excitations at the same cost having converged just the occupied KS states for the ground state.\cite{Bruner2016,Repisky2015} Besides, propagation of the KS states can be highly parallelized enabling to compute optical properties for up to several thousands of atoms. \cite{Jornet-Somoza2015, Andrade2012}.

These features make P-TDDFT the suitable theoretical framework to study photo-active complex systems such as natural LHC and OLEDS. In such cases, it is important to take into account the environment effects, which is commonly done by adding polarizable force fields\cite{Donati2017} or by other methods.\cite{Mennucci2011}

It is well known that the absorption cross-section can be computed from the propagation of the KS states in the linear-response regime. Besides, the transition densities and plasmons can be qualitatively evaluated from the Fourier transform of the time-dependent induced density.\cite{Thiele2009, Hofmann2012, Wang2012, Fischer2015}  Recently Schelter et at. \cite{Schelter2018} provided an analytical form to obtain the oscillator strengths, excitation energies and transition densities from the real time propagation of the electron system. They used the many-body ground state Hamiltonian to show that a quantum mechanical solution for the response functions after a boost excitation have the cardinal sine form and they exemplified their method by performing real-time propagation TDDFT calculations. In that paper, however, they did not address the way how transition densities can be extracted for randomly oriented distorted chromophores in complex systems.

In this work, we provide another derivation using the linear-response formalism, which makes possible to study exciton couplings of arbitrary complex systems.  We present a theoretical description of bright excitations by performing P-TDDFT that can be applicable even to very large systems, all treated entirely at the same atomistic level of theory. Combining the \textit{local density analysis} we recently introduced \cite{Jornet-Somoza2015} and our new derivation described below, the individual transition dipole moments and transition densities are obtained from P-TDDFT. The techniques we propose here permit to compute not just the first excitation but a broad energy range of excited states of molecules taking into account all effects induced by the rest of the system. This methodology provides a powerful tool to study exciton dynamics in light-harvesting complexes\cite{JornetSomoza2019} and light-emitting layers of OLEDs.\cite{Lebedeva2019}

 As a proof of concept of the method here presented, we illustrate its validity for the cases of increasing complexity including the benzaldehyde molecule, dimer and a cluster of 14 molecules. 
 In the following we first describe the methodology developed and then discuss the results for the systems listed.

\section{THEORETICAL DEVELOPMENT}\label{sec:Theory}
In this section we first describe the fundamentals of computation of  absorption spectra for finite systems from real-time propagation TDDFT (P-TDDFT). 
Then, we expose our time-dependent local-density analysis that enables to decompose the absorption spectra into individual contributions. 

In the next subsection,  we derive analytical expressions for the transition moment and transition densities for a given excitation from P-TDDFT in the linear-response regime. 

In the third subsection, both techniques are combined in order to study exciton transfer in multichromophore systems taking into account the actual environment, treated entirely at the same level of theory, i.e. TDDFT.

\subsection{Absorption Spectra Decomposition}

In the dipole approximation, the influence of the electric field applied to a quantum system can be supervised by the time evolution of the dipole moment:

\begin{equation}\label{eq:tddipole}
\mu_{\nu}(t) = \mu_{\nu}(t_0) + \int_{-\infty}^{+\infty} \alpha_{\nu\lambda}(t-t') \mathcal{E}_{\lambda} (t') dt',
\end{equation}

\noindent where $t_0$ is the initial time just before the application of the time dependent external field, $\mathcal{E}_{\lambda} (t)$, and the dynamic polarizability element, $\alpha_{\nu\lambda}(t-t')$,  is the retarded response that relates the $\lambda$ component of the external electric field to the change of the $\nu$ component of the dipole moment of the system. 

In the linear-response theory, the dynamic polarizablity tensor is constructed from the general expression for the retarded response function \cite{Ullrich2012}:
\begin{equation}
 \alpha_{\nu\lambda}(t-t') = -i\Theta(t-t')\langle \Psi_0 |[\hat \mu_{\nu}(t-t'), v_{\lambda}^{ext}]| \Psi_0\rangle,
\end{equation}

\noindent where the dipole moment operator and the external potential are described as $\hat \mu_{\nu} \color{black}= e\hat r_{\nu}$ and $\delta v_{\lambda}^{ext}(t)\color{black} = \hat r_{\lambda} \mathcal{E} (t)$, respectively (where $e$ refers to the electron charge), $\Theta(t-t')$ is the Heaviside function and $| \Psi_0\rangle$ is the the wave function at the initial moment of time ($t_0$).

The polarizability tensor $\boldsymbol{\alpha(\omega)}$ can be obtained by propagation of the KS states after the application of a perturbative boost. Usually, for finite systems, this external pertubative potential is applied in the form of a delta-pulse of the electric field \cite{Marques2012,Ullrich2012}:
\begin{equation}\label{eq:vextkick}
 v^{ext} = -e \textbf{r} \cdot \textbf{k} \delta(t_{0}),
\end{equation}
where $\textbf{k}$ corresponds to the magnitude vector of the electric field, which should be sufficiently small to ensure the linear-response regime.

Then, the polarizability tensor can be written in frequency space ($\omega$) in terms of the Fourier Transform of the time-dependent dipole moment as

\begin{equation}\label{eq:dynpoltensor}
\alpha_{\nu\lambda}(\omega) = \frac{1}{k_\lambda} \int_0 ^{\infty} dt [\mu_\nu(t) - \mu_\nu(t_{0})]e^{-i\omega t}.
\end{equation}
\noindent Note that if $\textbf{k}$ is sufficiently small so that the system is in the linear-response regime, the polarizability tensor does not depend on it.

In the time-dependent Kohn-Sham scheme, the time-dependent dipole moments can be easily obtained from the time-dependent electronic density ($\rho(\boldsymbol{r},t))$):

\begin{equation}\label{eq:tm_from_tp}
 \mu(t) = \int \textbf{r} \cdot \rho(\boldsymbol{r},t) d\textbf{r},
\end{equation}
where
\begin{equation}
 \rho(\boldsymbol{r},t) = \sum_{i=0}^{N_{occ}} |\Phi_i(\boldsymbol{r},t)|^2,
\end{equation}
where $\Phi_i(\boldsymbol{r},t)$ are the KS orbitals and the summation runs over the $N_{occ}$ occupied orbitals. 

At the same time, by describing the dynamic polarizability tensor in the Lehmann's representation and using the fluctuation-dissipation theorem,\cite{Ullrich2012} the imaginary part of the dynamic polarizability tensor can be related with the oscillator strength of each transition, in atomic units, as

\begin{equation}\label{eq:spectralfunction}
  \frac{1}{3}\text{Tr}(\Im m\boldsymbol \alpha (\omega)) = \frac{\pi}{3}\sum_{n=1}^\infty \sum_{\nu = 1}^{3}  |\langle \Psi_{0} |\hat r_\nu| \Psi_{n}\rangle |^2 \delta(\omega-\Omega_{n})=\sum_{n=1}^\infty  \frac{\pi}{2\Omega_{n}}f_{n}\delta(\omega-\Omega_{n}),
\end{equation}

\noindent where the $| \Psi_{n} \rangle$ is the wave function of the $n$-th excited state, $\Omega_{n}$ is the corresponding excitation energy and $f_n$ is the oscillator strength which gives the transition probability for that particular excitation. This expression relates directly the dynamic polarization tensor ($\boldsymbol{\alpha}(\omega)$) and the transition dipole moment ($\vec{\mu}_{n0} = \langle \Psi_{0} | \hat{\boldsymbol{r}} | \Psi_{n} \rangle$). 

In addition, from the imaginary part of the dynamic polarizability tensor we can obtain the photo-absorption cross-section, which is gives the absorption spectrum of the system:

\begin{equation}\label{eq:photo-absorption_spectrum}
    \sigma(\omega)=\frac{4\pi\omega}{3c}Tr[\Im m(\boldsymbol{\alpha}(\omega)] = \frac{2\pi^{2}}{c}S(\omega)
\end{equation}

\noindent where $c$ is the speed of light and $S(\omega)= \sum_{n=1}^\infty f_n\delta(\omega-\Omega_n)$ is the dipole spectral function.

\subsection{Local Density Analysis for Spectrum Decomposition}
The first time we introduced this methodology was in the decomposition of theoretical spectra of the major light harvesting complex II (LHCII) \cite{Jornet-Somoza2015}. This procedure based on the DFT fundamentals, in particular on the DFT theorem exposed by Hohenberg and Kohn, that establish the electron density as a basic variable from which any observable property of a quantum system can be calculated.\cite{Hohenberg964}  Then, we can define the local observable property by splitting the total electronic density into the contributions for different subsystems. This partitioning is performed using the quantum theory of atoms in molecules (QTAIM) introduced by Richard Bader.\cite{Bader1994} Among other topological properties, QTAIM enables to assign different regions of space to specific atoms by following the gradient of the electron density. 

Therefore, we can decompose our total electron density into the sum of electron densities which belong to different molecules forming the complex system of interest. 
Using this approach, we can determine the local time-dependent density for each chromophore and compute its contribution to the photo-absorption spectrum by applying eqs \ref{eq:tm_from_tp}-\ref{eq:photo-absorption_spectrum}.

It is important to remark that, since we stay in the linear-response regime, the boost on the initial wave function does not produce a significant change in the electron density distribution. Instead just small fluctuations are observed. For this reason, partitioning of the ground state density is enough to define individual molecular space regions, and to study the time evolution of the electron density inside these domains.

\subsection{Transition Dipole Moments and Transition Densities from P-TDDFT}

In order to properly characterize electronic transitions it is worthy to obtain transition dipole moments, $ \Vec{\mu}_{n0} = \langle \Psi_0 |\hat{\boldsymbol{r}}| \Psi_n\rangle$,  and transition densities, $\rho^{T}_{I0}(\boldsymbol{r})=\langle \Psi_I |\hat{n}(\boldsymbol{r})| \Psi_0\rangle$, where $\hat{\boldsymbol{r}}$ is the position operator and $\hat{n}(\boldsymbol{r})$ the density operator. From eqs \ref{eq:tm_from_tp}-\ref{eq:spectralfunction} we can see that transition dipole moments are related with the imaginary parts of Fourier transforms of time-dependent dipole moments. Analogously, in P-TDDFT,  transition densities are related with the imaginary parts of  Fourier transforms of induced time-dependent densities. \cite{Kummel2001,Thiele2009,Hofmann2012,Wang2012}.

Recently, Schelter et at.\cite{Schelter2018} derived an analytic expression to compute transition dipole moments and transition densities from P-TDDFT calculations, based on the coefficient expansion of the time-dependent wave function in terms of eigenfunctions of the many-body ground-state Hamiltonian.  

In this work, we present a complementary derivation of both properties based only in LR-TDDFT arguments. 
Let us start with the key object in the linear-response TDDFT, which is the definition of the response density as
\begin{equation}\label{eq:response_density}
\delta\rho(\boldsymbol{r},t) = \int_{t_0}^{t}dt'\int d\boldsymbol{r}' \chi_{nn'}(\boldsymbol{r}t,\boldsymbol{r}'t') v^{ext}(\boldsymbol{r}',t'),
\end{equation}

\noindent where $\chi_{nn'}$ is the density-density response function: \cite{Marques2012} 
\begin{equation}\label{eq:nn-response}
\chi_{nn'}(\boldsymbol{r}t,\boldsymbol{r}'t') = -i\theta (t-t') \langle \Psi_0 |[\hat{n}_{H_0}(\boldsymbol{r},t), \hat{n}_{H_0}(\boldsymbol{r}',t')]| \Psi_0\rangle. 
\end{equation} 

\noindent The density operator here is defined in second quantization as $\hat{n}=\hat a_j^{\dagger} \hat a_i$, and the subindex $H_0$ stands for the zero-order Hamiltonian, i.e. with no external pertubation. 

In the frequency space this gives the well-known "Lehmann-representation": 
\begin{equation}\label{eq:lehmann}
\chi_{nn'}(\boldsymbol{r},\boldsymbol{r}', \omega) = \sum_{I\neq 0} \Bigg \{ \frac{ \langle \Psi_0 |\hat{n}(\boldsymbol{r})| \Psi_I\rangle \langle \Psi_I |\hat{n}(\boldsymbol{r}')| \Psi_0\rangle}{\omega - \Omega_I +i\eta}  
- \frac{\langle \Psi_0 |\hat{n}(\boldsymbol{r}')| \Psi_I\rangle \langle \Psi_I |\hat{n}(\boldsymbol{r})| \Psi_0\rangle}{\omega + \Omega_I +i\eta} \Bigg \}. 
\end{equation}

From this expression we can see that the information regarding all transition densities for excitations from the ground state ($\Psi_0$) to any excited state ($\Psi_I$) is kept in the density-density response function, since the transition density is defined by $\rho^{T}_{I0}(\boldsymbol{r})=\langle \Psi_I |\hat{n}(\boldsymbol{r})| \Psi_0\rangle$.

We can now make use of the "fluctuation-dissipation theorem" \cite{Ullrich2012}, that relates the density-density response function ($\chi_{nn}$) with the corresponding dynamical structure factor ($S_{nn}$) at zero temperature:

\begin{equation}\label{eq:chi-s}
\Im m\chi_{nn}(\boldsymbol{r},\boldsymbol{r}', \omega) = -\pi\big(S_{nn}(\boldsymbol{r},\boldsymbol{r}', \omega)-S_{nn}(\boldsymbol{r}',\boldsymbol{r}, -\omega)\big )
\end{equation}
\begin{equation}\label{eq:snn}
S_{nn}(\boldsymbol{r},\boldsymbol{r}', \omega) = n_0(\boldsymbol{r})n_0(\boldsymbol{r'})\delta(\omega) + \sum_{n=1}^\infty \langle \Psi_0 | \hat{n}(\boldsymbol{r}) | \Psi_n \rangle \langle \Psi_0 | \hat{n}(\boldsymbol{r'}) | \Psi_n \rangle\delta(\omega-\Omega_n).
\end{equation}

 Since we assume $T=0$ K , the dynamical structure factor fulfills $S_{nn}(\omega)=0$ for $\omega < 0$ and $S_{nn}(-\omega)=0$ for $\omega > 0$. Then, we do not consider \textit{stimulated emission} and focus on the absorption spectrum (i.e. consider just $\omega > 0$). By insterting eq \ref{eq:chi-s} in eq \ref{eq:response_density}, where the system is perturbed by an instantaneous electric field, and using eq \ref{eq:vextkick}, we obtain the following equation:
\begin{equation}
\Im m\delta\rho(\boldsymbol{r},\omega) = \int d\boldsymbol{r'} \pi S_{nn}(\boldsymbol{r},\boldsymbol{r}', \omega) e \boldsymbol{r'}\cdot\boldsymbol{k} \delta(\omega-\Omega_n).
\end{equation}

Taking into account that the component of the transition moment of a given excitation can be computed from the corresponding transition density (i.e. $\mu_{n,\nu}= \int r_\nu \rho_{n0}^{T}(\boldsymbol{r})d\boldsymbol{r}$) , one gets
\begin{equation}
\int  \boldsymbol{r}\cdot \boldsymbol{k} \langle \Psi_0 | \hat{n}(\boldsymbol{r}) | \Psi_n \rangle   d\boldsymbol{r} = \mu_{n,k} \cdot k,
\end{equation}
where $k$ is the magnitude of the external electric field (eq \ref{eq:vextkick}), and $\mu_{n,k}$ corresponds to the projection of the transition dipole moment ($0\rightarrow n$, $n$ excited state) over the polarization direction of the electric field. 

Then, the imaginary part of the response density can be expressed in terms of all transition densities scaled by the projection of the corresponding transition moment on the direction of the external field, 

\begin{equation}
\Im m\delta\rho (\boldsymbol{r}, \omega) = \pi k \sum_{n=1}^\infty \mu_{n,k} \langle \Psi_0 | \hat{n}(\boldsymbol{r}) | \Psi_n \rangle \delta(\omega-\Omega_n). 
\end{equation}

If we are interested in the excited states that are energetically far one from each other, the transition densities can be recovered from the response densities under an active perturbation, i.e. the one for which the transition is dipole allowed and $\mu_{n,k} \neq 0$,

\begin{equation}\label{eq:lrtransdens}
\langle \Psi_0 | \hat{n}(\boldsymbol{r}) | \Psi_n \rangle \delta(\omega-\Omega_n) = \frac{\Im m\delta\rho (\boldsymbol{r}, \omega)}{\pi k \mu_{n,k}}.
\end{equation}

\subsubsection{Finite propagation times in P-TDDFT}

As stated in eq \ref{eq:lrtransdens}, we first need to know the transition moment for a given excitation in order to recover the transition density. 

To obtain the photo-absorption spectrum from finite propagation times in  P-TDDFT, a damping function is commonly added in the Fourier transform of the polarizability tensor determined by eq \ref{eq:dynpoltensor} (see Supplementary Information). It induces an artificial decay of the excited population and allows to get a smooth spectrum removing artificial peaks. In particular, in our calculations we introduce a Gaussian damping function ($e^{-\eta^2 t^2}$). We set the parameter $\eta$ such that the damping function reaches the value of $1e^{-4}$ at the end of the propagation time $\tau$: 

\begin{subequations}
\begin{equation}
D(t) = e^{-\eta^2 t^2},
\end{equation}
\begin{equation}
\eta = \frac{\sqrt[]{-ln (1e^{-4})}}{\tau}.
\end{equation}
\end{subequations}

\noindent The damping parameter is responsible for the spectral broadening (see Supplementary Information) so that each single excitation is represented by a Gaussian function with half-width at half-maxima ($\sigma_{hwhm}$)

\begin{equation}
\sigma_{hwhm} = 2\sqrt[]{ln2}\cdot\eta,
\end{equation}

Transition dipole moments can be obtained from a P-TDDFT calculation by diagonalization of the dynamic polarizability tensor given by eq \ref{eq:dynpoltensor}. The frequency-dependent diagonal elements of the diagonalized polarizability tensor give the information about the probability of orthogonal excitations (i.e. transitions charactherized by orthogonal transition dipoles), and enables to distinguish between quasi-degenerate states.

The area under the trace of the diagonalized dynamic polarizability tensor is proportional to the square of the transition dipole moment (see eq \ref{eq:spectralfunction}) and the direction of the transition dipole moment is determined by the corresponding eigenvectors. 

In order to obtain the magnitudes of the transition dipole moments, we fit the spectra generated from the diagonalized frequency-dependent dynamic polarizability tensor by a sum of Gaussian functions. The Gaussian fit is obtained using the methodology described by Goshtasby et al. \cite{Goshtasby1994}.
Then the broadening of the Fourier Transform and the location of the excitation energies are found from the fitted parameters of the Gaussian half-width at half-maxima ($\sigma_{n}$) and the Gaussian center, respectively ($\epsilon_{n}$). 

Knowing the transition dipole moments and broadening introduced due to the damping function, transition densities can be computed from P-TDDFT on finite times by taking the imaginary part of the Fourier transform of the time-dependent response density applying the same Gaussian damping. Then, integrating eq \ref{eq:lrtransdens} around the peak at $\Omega_n$, we obtain the analytical expression for the transition density:

\begin{equation}\label{eq:n0j-PTDFT}
\rho_{n0}^{T}(\boldsymbol{r})=\langle \Psi_0 | \hat{n}(\boldsymbol{r}) | \Psi_n \rangle  = \frac {\sigma_n}{\mu_{n,k} k\sqrt{\pi\ln 2} } \Im m\delta\rho (\boldsymbol{r}, \Omega_n),
\end{equation}

\noindent where $\sigma_n$ is precisely the Gaussian fitted half-width at half-maxima around $\Omega_n$.

\subsection{Exciton Coupling Evaluation}
Both techniques, the local density analysis and calculation of transition densities from P-TDDFT, can be combined in order to study complex systems containing more than one photo-active molecule.

The exciton dynamics is governed by the exciton coupling determined by the Frenkel Exciton Hamiltonian: 
\begin{equation}\label{eq:frenkelh}
\hat{H} = \sum_n \epsilon_n|n\rangle\langle n| + \sum_{n,m}V_{nm}|m\rangle\langle n|,
\end{equation}

\noindent where $|n\rangle$ is the excited state located on the site $n$, $\epsilon_n$ are the site energies (or vertical excitation energies) and $V_{nm}$ corresponds to the \textit{exciton coupling} between excited states on the donor site $n$, and the acceptor site $m$. The \textit{exciton coupling} is determined by the interaction between the transition densities, \cite{Hsu2001,Curutchet2017}

\begin{equation}\label{eq:Vnm}
V_{nm} = \int d\textbf{r} \int d\textbf{r}' \rho^{T*}_m(\textbf{r}')\left(\frac{1}{|\textbf{r}'-\textbf{r}|} + g_{xc}(\textbf{r}',\textbf{r})\right)\rho^{T}_n(\textbf{r}) - \omega_{n/m} \int d\textbf{r}\rho^{T*}_m(\textbf{r})\rho^{T}_n(\textbf{r}),
\end{equation}

\noindent  where $g_{xc}$ is the exchange-correlation kernel, and $\omega_{n/m}$ is the resonant excitation energy for states $n$ and $m$ for symmetric systems, otherwise the average of the acceptor ($\omega_n$) and donor ($\omega_m$) excitation frequencies is used. Note that this expression was derived considering the interaction between chromophores as a perturbation. It is not accurate when the chromophores are so close that their electronic wave functions strongly interact, mostly due to neglect of the charge-transfer effects.

\begin{figure}
    \centering
    \begin{subfigure}{.475\textwidth}
       \centering
       \includegraphics[width=0.7\textwidth]{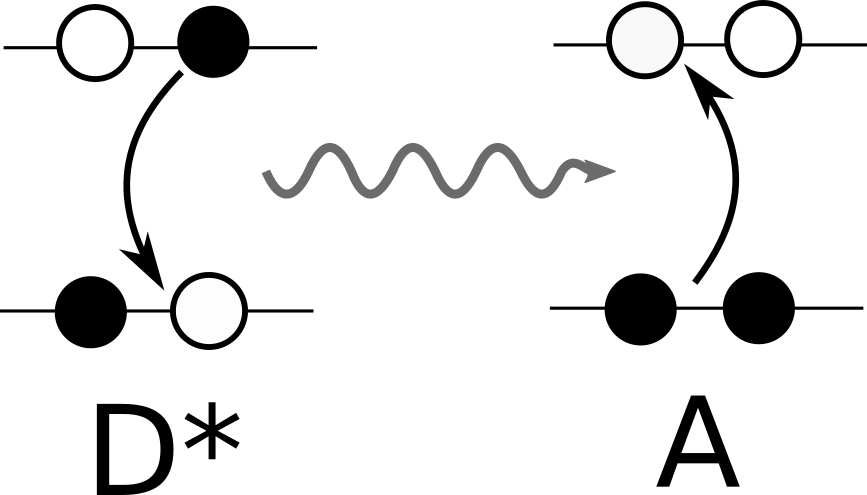}
       \caption{}
       \label{fig:forster}
    \end{subfigure}
    \begin{subfigure}{.475\textwidth}
       \centering
       \includegraphics[width=0.7\textwidth]{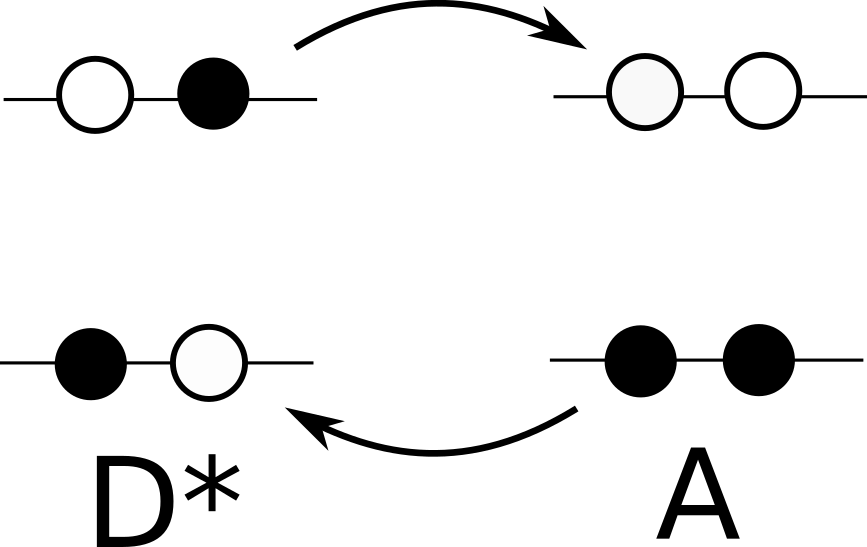}
       \caption{}
       \label{fig:dexter}
    \end{subfigure}
    \caption{Schematic representation of the excitation energy transfer for (a) F\"orster and (b) Dexter mechanisms.}
    \label{fig:eetmech}
\end{figure}

The Coulombic term inside the kernel function is usually attributed to the F\"orster resonant energy transfer (FRET) \cite{Forster1946}. This term stands for the interchange of the exciton population between two sites where the fluorescent energy of the donor is absorbed by the acceptor producing a resonant excitation. The second term in the kernel function corresponds to the exchange-correlation effect, and is related the Dexter electron transfer.\cite{Dexter1953} In this mechanism, the orbital proximity enables the simultaneous interchange of (i) an excited electron from the donor to an unoccupied orbital of the acceptor, and (ii) an electron from highest occupied molecular orbital (HOMO) of the acceptor to a hole on low lying single occupied molecular orbital (SOMO) of the excited donor (Figure \ref{fig:eetmech}). 

There are several ways to compute the \textit{exciton coupling} between two excited states of different sites.
The most commonly used ones are the \textit{ideal dipole approximation} (IDA)\cite{Andrews1989} and the \textit{transition density cube} (TDC) method.\cite{Krueger1998} The former assumes that both, acceptor and donor molecules, are far enough with no overlap of the electronic density and then the Coulomb interaction can be written in terms of the point dipole approximation as
\begin{equation}\label{eq:fret}
V^{FRET}_{IDA} = \kappa^{DA} \frac{|\boldsymbol\mu^A||\boldsymbol\mu^D|}{|\textbf{R}|^3}
\end{equation}

\noindent with the orientation factor $\kappa^{DA} = \boldsymbol{u}^D \cdot \boldsymbol{u}^A - 3(\boldsymbol{u}^D \cdot \boldsymbol{u}^R)(\boldsymbol{u}^A \cdot \boldsymbol{u}^R)$, where $\boldsymbol{u}^{D,A}$ are the unit vectors describing directions of transition dipole moments for the donor (D) and acceptor (A), $\boldsymbol{R}$ is the distance vector connecting both centers of mass, and $\boldsymbol{u}^{R}$ is the unit vector along the same direction.  

This approximation does not take into account the delocalization of the transition densities over the molecules. In other words, different regions of the electronic density of the acceptor molecule feel different potential generated by the excited density of the donor molecule due to the spacial arrangement. 

The use of the IDA approximation requires the knowledge of the transition dipole moments both for donor and acceptor molecules. In the Section \ref{sec:Applications} we 
compare exciton couplings computed using different transition dipole moments obtained  i) from consideration of isolated molecules; named here as $V(IM)$ (where \textit{IM} stands for Isolated Molecule) and ii) from the proposed \textit{local density analysis}, $V(LDT)$ (where \textit{LDT} stands for local density treatment, not to confuse with the LDA exchange-correlation functional).  

In the TDC method, the transition densities are discretized into small volume units and then the exciton coupling is computed through the following sum over the volume units:

\begin{subequations}\label{eq:tdcm}
  \begin{equation}
    M^{n0}_X(x,y,z)=V_\delta\int_z^{z+\delta z}\int_y^{y+\delta y}\int_x^{x+\delta   x}\Psi_0^X\Psi_n^{X*}\text{d}x\text{d}y\text{d}z
  \end{equation}
  \begin{equation}
    V_s = \int d\textbf{r} \int d\textbf{r}' M^{n0}_A(\textbf{r}'_A)\left(\frac{1}{|\textbf{r}'-\textbf{r}|} + g_{xc}(\textbf{r}',\textbf{r}) \right)M^{0n}_D(\textbf{r}_B) 
  \end{equation}
\end{subequations}

The TDC method can therefore include both Coulomb and exchange-correlation contributions to the exciton coupling (eq \ref{eq:Vnm}) by evaluating the Coulomb and exchange-correlation potentials generated by the donor molecule in the region embracing the acceptor molecule. The exciton coupling obtained using the TDC method with the local transition densities calculated within our P-TDDFT formalism is hereafter denoted as $V(TDCM)$.

\subsection{Procedure and Computational Details}\label{sec:ComputDetails}

\begin{figure}[!ht]
    \centering
    \includegraphics[width=0.75\textwidth]{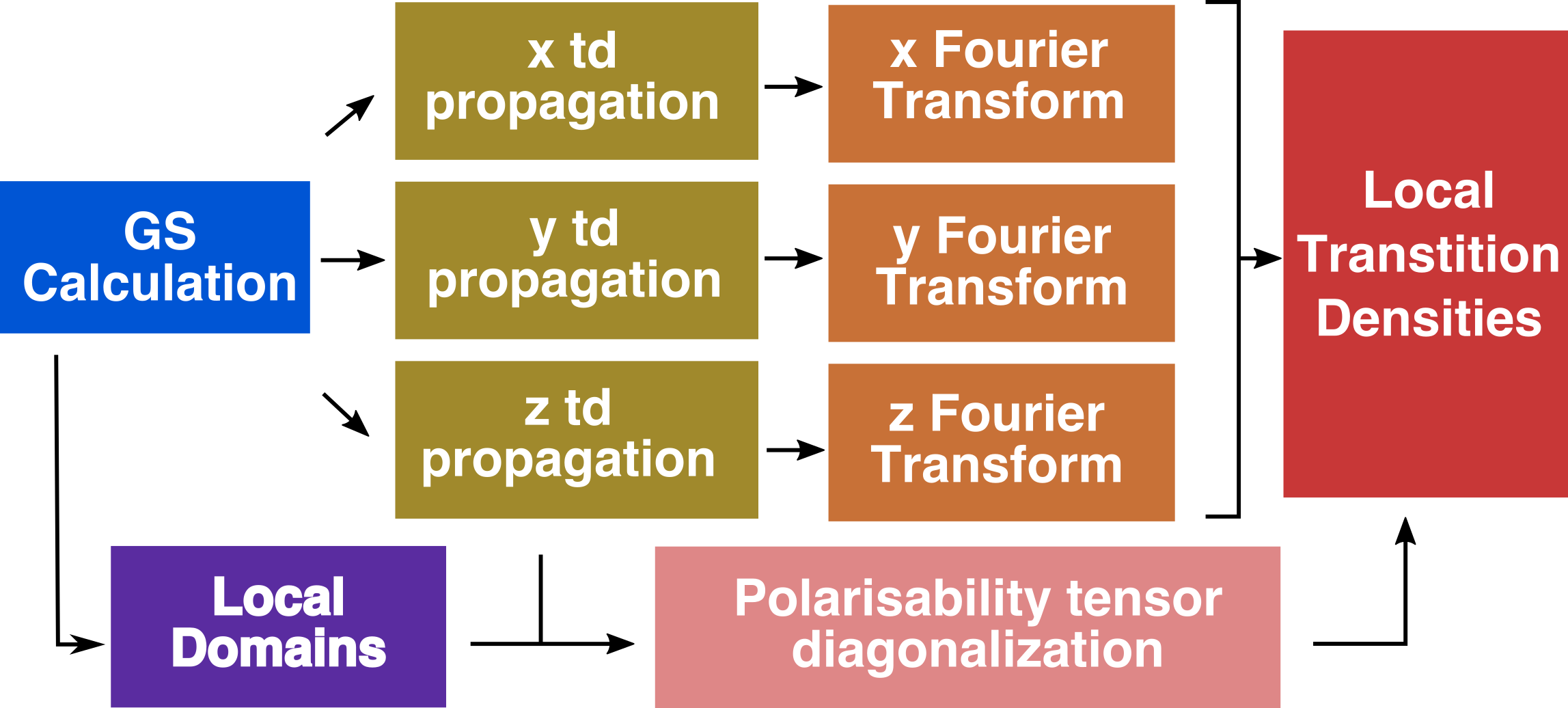}
    \caption{Scheme of the procedure used to get transition dipole moments and transition densities from P-TDDFT calculations for multichromophore systems.}
    \label{fig:ld-scheme}
\end{figure}

Summarizing,  local transition densities for a complex multichromophore system can be obtained the following computational steps shown in Figure \ref{fig:ld-scheme}. After preforming a ground state (GS) calculation, the local domains are determined to get the densities belonging to each molecule. We have checked that, since we stay in a linear-response regime during the time dependent  (td) propagation, the electronic charge inside each domain does not vary significantly (less than 0.005\%, see Figure \sref{SIfig:ld_gsvstd})
), hence there is no need to update the grid point assignment to different spatial domains during the time-propagation calculations. Three P-TDDFT calculation are executed by perturbing the initial state with an electric field polarized in each of the Cartesian directions, recording the dipole moments and saving the time-dependent density every 20 steps of the propagation. Then the dynamic polarizability tensor for each domain is obtained and diagonalized to get the local contribution to the absorption spectrum as well as the transition dipole moments. The frequency-dependent response densities are obtained by the Fourier Transform of the stored time-dependent electron density. Finally the transition densities for each molecule and excitation can be reconstructed and the exciton coupling can be computed using the TDC method. 

\texttt{OCTOPUS} is a quantum chemistry/physics code specially designed to solve ground-state  and time-dependent (TD) DFT problems.\cite{Marques2003,Castro2006,Andrade2015}   In \texttt{OCTOPUS} the basic quantities (Hamiltonian, potentials and single-particle wave functions) are considered on a discretized real-space grid. Besides, \texttt{OCTOPUS} is highly parallelized in both grid points and Kohn-Sham states, which  enables fast evaluation of non-linear propagation equations for the initial Slater determinant during real-time electron dynamics. These features make \texttt{OCTOPUS} the perfect code to implement our derivation of transition densities and exciton couplings from P-TDDFT. 

In this work we also use the ORCA package that allows us efficient computation of excited state properties by solving the Casida's equation using atom localized basis sets.\cite{Neese2012,Neese2018} ORCA package uses several approaches to reduce the computational cost providing a good compromise against the accuracy of the results, e.g. the resolution of the identity (RI) approximation for the Coulomb potential.\cite{Eichkorn1995}

In order to be sure that both types of the calculations (using ORCA and \texttt{OCTOPUS}) are consistent, we compared the ground-state eigenvalues (see Table \sref{SItbl:ks_oct_vs_orca}) and excited-state energies (see Table \ref{tbl:casida_excitations}). It can be seen that in all cases the energy difference between the methods is smaller than 0.05 eV.

Further computational details of the geometry optimization, ground-state calculations and TDDFT calculations using linear-response and time-propagation schemes are described in Supplementary Information.  The calculated numerical data that support our study are available in "NOMAD repository" ( http://dx.doi.org/10.17172/NOMAD/2019.02.27-2).

\section{Applications and Discussion}\label{sec:Applications}
\subsection{Isolated Molecule}
\begin{figure}[h!]
  \centering
  \begin{subfigure}{.49\textwidth}
     \includegraphics[width=0.75\textwidth]{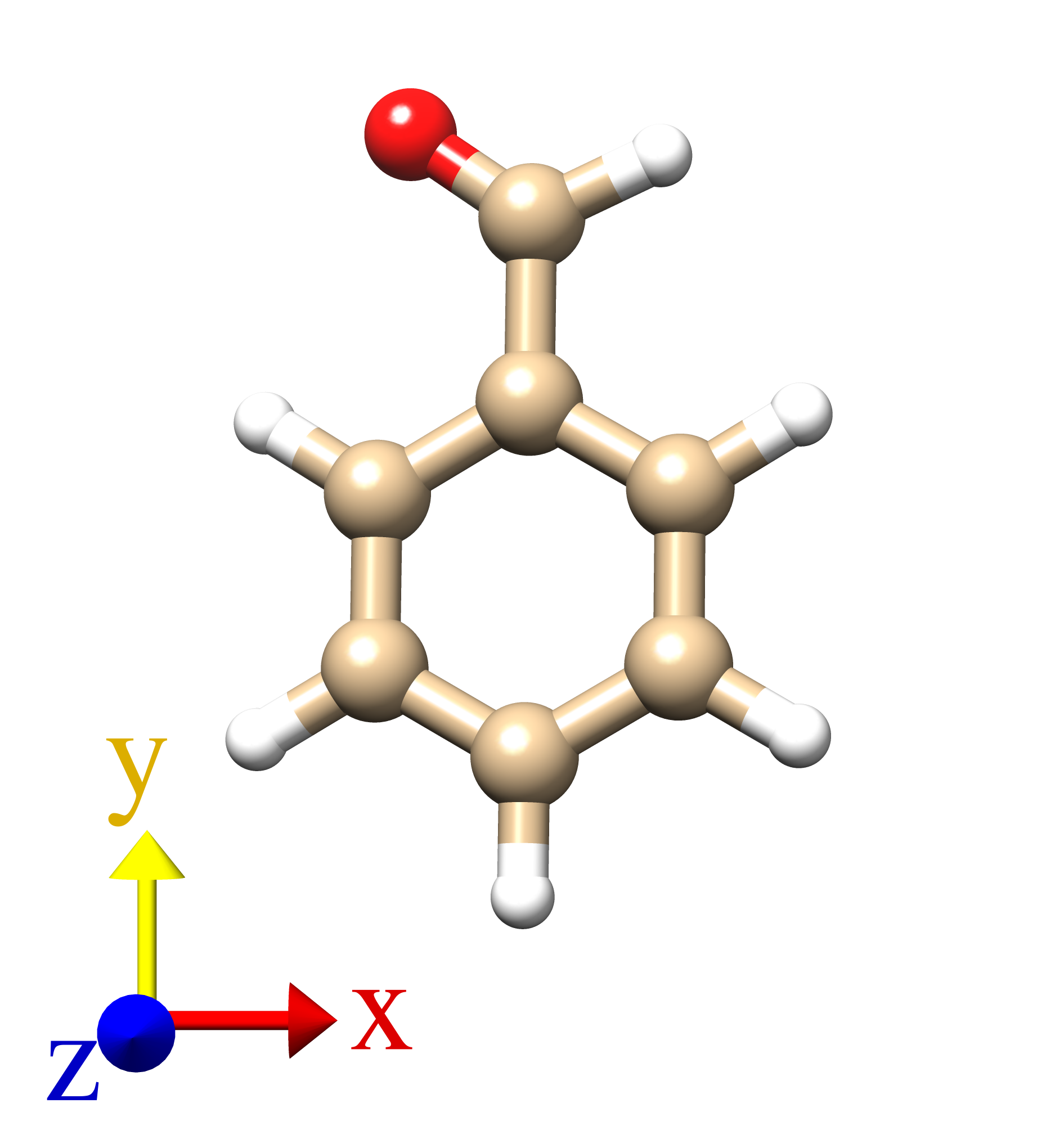}
     \caption{}
     \label{fig:molecule}
  \end{subfigure}
  \begin{subfigure}{.49\textwidth}
     \includegraphics[width=1.\textwidth]{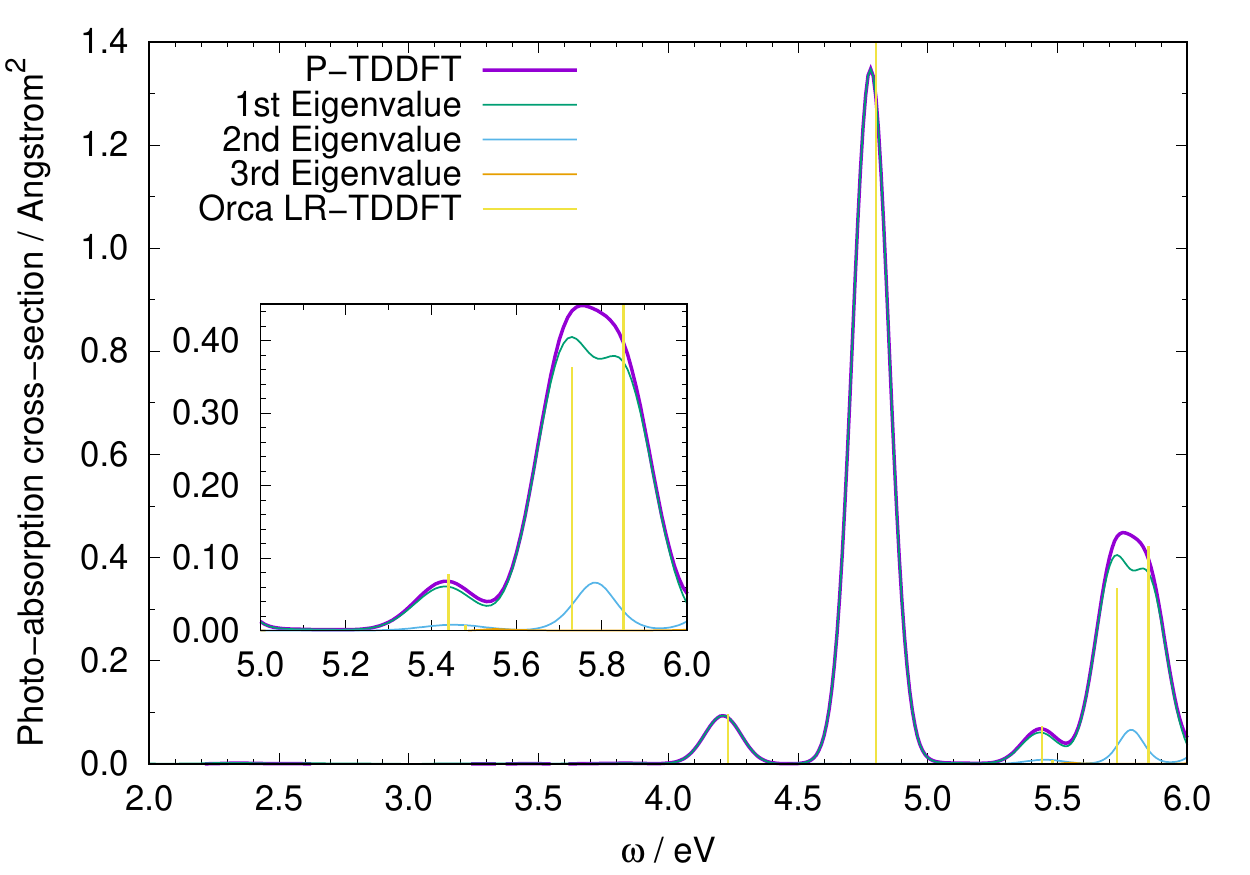}
     \caption{}
     \label{fig:casida_spectra}
  \end{subfigure}
  \caption{(a) Benzaldehyde molecule orientation. Brown, white and red spheres represent carbon, hydrogen and oxigen atoms, respectively. (b) Absorption spectrum of the isolated benzaldehyde  molecule computed using P-TDDFT (purple line) and LR-TDDFT (yellow impulses). The decomposition of P-TDDFT results into the contributions from different eigenvalues of the diagonalized polarizability tensor indicated using green, blue and orange lines.}
  \label{fig:bnzl}
\end{figure}

Let us exemplify the applicability of the formalism described above by computing the excitation properties for a single benzaldehyde molecule (Figure \ref{fig:molecule}). Figure  \ref{fig:casida_spectra} shows the benzaldehyde UV-Visible absorption spectrum computed solving the Casida's equation within linear-response TDDFT.\cite{Casida1995} It presents two major optically allowed excitations at 4.2 eV and 4.8 eV.

\begin{table}[h!]
\centering
\begin{tabular}{ c | c | c | c | c | c }
 Method &  $\epsilon_n$ & $\mu_{x}$  & $\mu_{y}$  & $\mu_{z}$   & $f_{n}$ \\
  \hline
  P-TDDFT & 4.21       &  0.1972 & 0.0055 & 0.0439 &  0.0150 \\
  LR-TDDFT & \textit{4.23}       &  \textit{0.1982} & \textit{0.0042} & \textit{0.0440} &  \textit{0.0152} \\
  P-TDDFT & 4.78       & -0.0091 & -0.7153 & 0.0914  & 0.2175 \\
  LR-TDDFT &\textit{4.80} & \textit{-0.0085} & \textit{-0.7178} & \textit{0.0915}   & \textit{0.2200} \\
\end{tabular}
\caption{Excitation energies ($\epsilon_{n}$, in eV), components of the transition dipole moment ($\mu_{x}, \mu_{y}, \mu_{z}$, in \AA) and oscillator strength ($f_{n}$) for the lowest bright excited states of the isolated benzaldehyde molecule determined by Gaussian fitting of the first eigenvalue of the diagonalized polarizability tensor. } 
\label{tbl:casida_excitations}
\end{table}

Figure \ref{fig:casida_spectra} also shows how the absorption spectra obtained from the trace of the dynamic polarizability tensor can be decomposed by performing a respective diagonalization. In the frequency regions where just single-degenerate bright states are present [4--5.5 eV], we can see that the first eigenvalue of this diagonalization matches to the total contribution of the spectrum. Whereas in the case when almost degenerate states are present [5.5--6 eV], we observe that although the trace of the dynamic polarizability shows just one broad peak, the first eigenvalue splits this band, in good agreement with the LR-TDDFT results. 

In Table \ref{tbl:casida_excitations} we give the values for the excitation energy and transition dipole moments obtained by the Gaussian fitting procedure of the dynamical polarizability tensor in the framework of P-TDDFT as implemented in the \texttt{OCTOPUS} package. \cite{Castro2006,Andrade2012} The values obtained from the solution of the Casida's equation as implemented in the ORCA package\cite{Neese2012} are also shown in rows 2 and 4. We can see that both methods give the same results even though they rely on very different approaches and implementations, validating in this way the methodology for calculation of transition dipole moments from P-TDDFT. 

\begin{figure}[h!]
  \centering
  \begin{subfigure}{.25\textwidth}
  \centering
  \includegraphics[width=.9\linewidth]{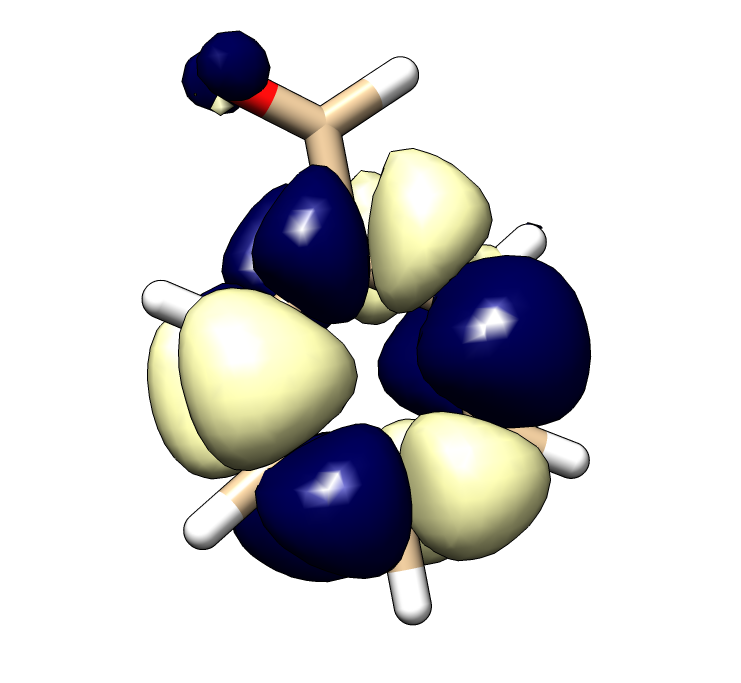}
  \caption{}
  \label{fig:td1}
  \end{subfigure}%
  \begin{subfigure}{.25\textwidth}
  \centering
  \includegraphics[width=.9\linewidth]{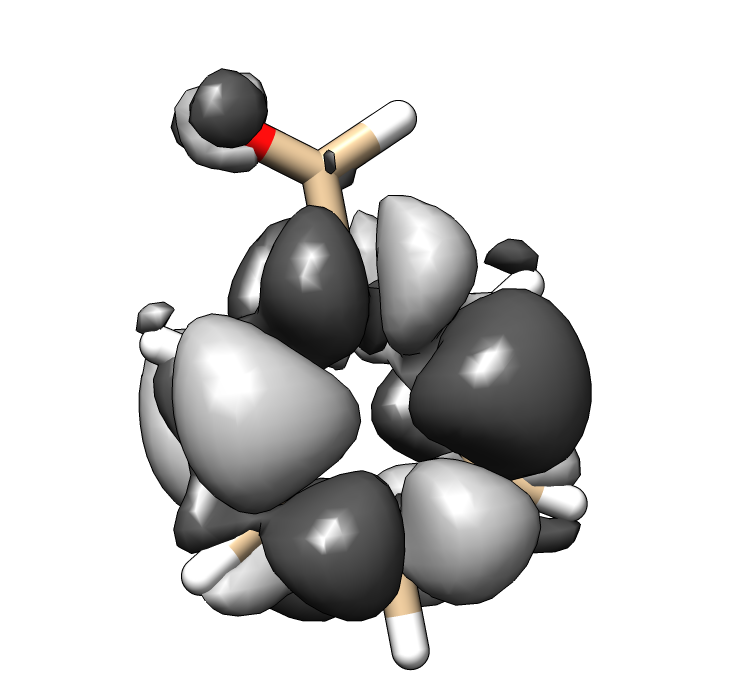}
  \caption{}
  \label{fig:td2}
  \end{subfigure}%
  \begin{subfigure}{.25\textwidth}
  \centering
  \includegraphics[width=.9\linewidth]{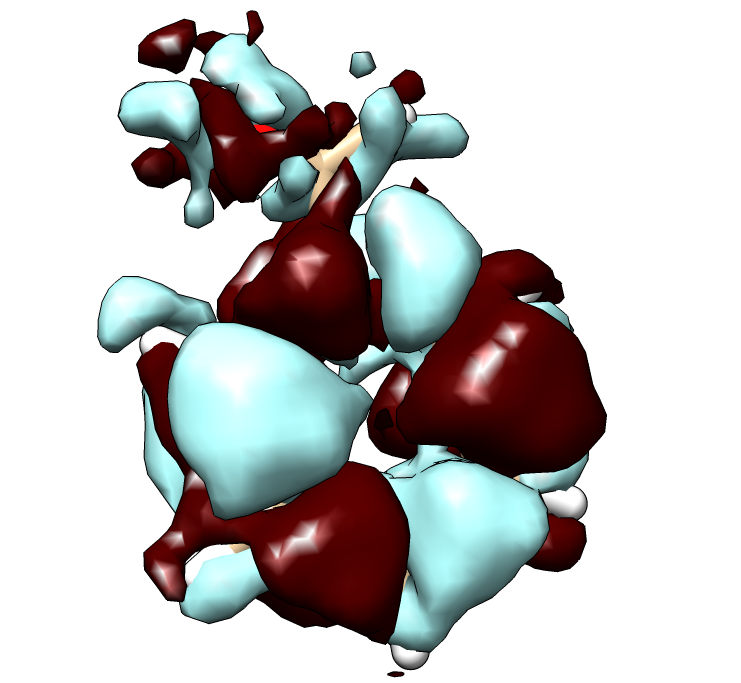}
  \caption{}
  \label{fig:td3}
  \end{subfigure}%
  \\
  \begin{subfigure}{.25\textwidth}
  \centering
  \includegraphics[width=.9\linewidth]{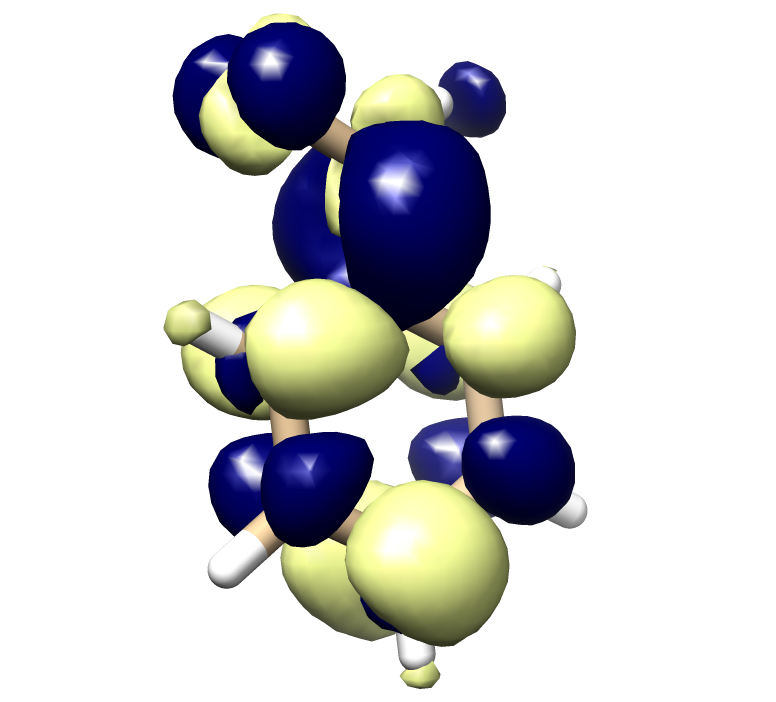}
  \caption{}
  \label{fig:td5}
  \end{subfigure}%
  \begin{subfigure}{.25\textwidth}
  \centering
  \includegraphics[width=.9\linewidth]{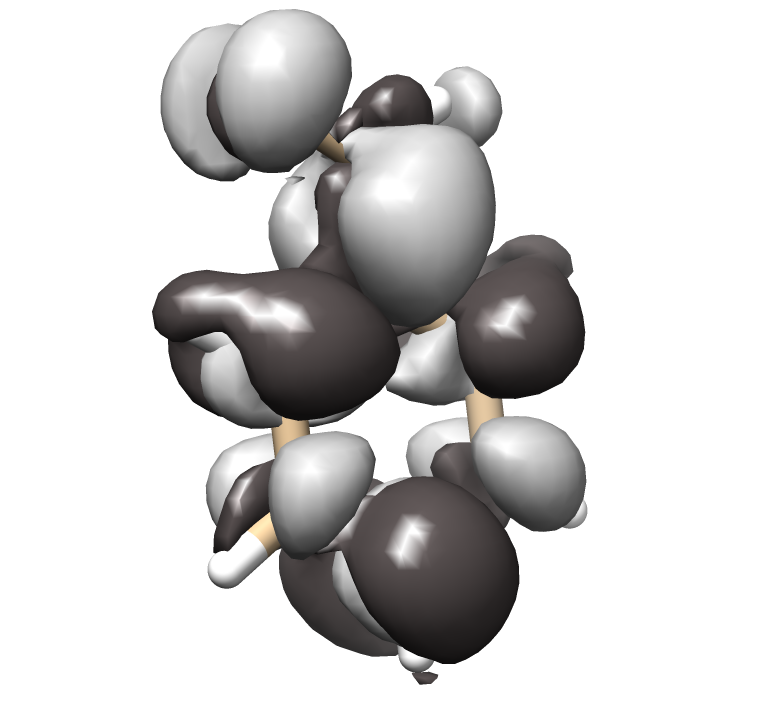}
  \caption{}
  \label{fig:td6}
  \end{subfigure}%
  \caption{Up: Transition densities (isovalue of 0.005 \AA\textsuperscript{-3}) for the first bright state of the isolated benzaldehyde molecule located at 4.2 eV obtained from (a) LR-TDDFT, (b) P-TDDFT with the electric field along the x axis, which corresponds to the maximum component of the transition dipole moment, and (c) average over P-TDDFT runs with the electric field along x, y and z axes. Bottom: Transition densities (isovalue of 0.0075 \AA\textsuperscript{-3}) for the second bright state located at 4.8 eV obtained from (d) LR-TDDFT and (e) P-TDDFT with the electric field along the y axis, which corresponds to the maximum component of the transition dipole moment.}
  \label{fig:transition_densities}
\end{figure}

Analogously, we can compare the respective transition densities obtained using both methods. Figure \ref{fig:transition_densities} shows the isocontours for the two bright states in Table \ref{tbl:casida_excitations}. The first transition density (Figure \ref{fig:td1}) is obtained using the eigenvectors of the Casida's equation. The transition densities in Figures \ref{fig:td2} and \ref{fig:td3} are obtained from the time-propagation of the Kohn-Sham orbitals after applying an initial perturbation keeping the system in the linear regime (P-TDDFT). From eq \ref{eq:n0j-PTDFT} it can be seen that we can obtain the transition density from any propagation where the projection of the transition dipole to the polarization direction of the electric field does not vanish. For the first excitation of the benzaldehyde molecule, the transition dipole moment is almost aligned along the x axis (Table \ref{tbl:casida_excitations} and Figure \ref{fig:molecule}). Though in principle, the components of the transition dipole moment for the other directions are not exactly zero, they take small values and their use for the calculations of the transition density can lead to errors. To demonstrate this we compared transition densities for the propagation run with the electric field along the x axis ($\boldsymbol{k}=(1, 0, 0)$) (Figure \ref{fig:td2}) and  the average over the propagation runs for all three axes x,y and z (Figure \ref{fig:td3}).  
We can see that the propagation with the electric field along the y axis ($\boldsymbol{k}=(0, 1, 0)$) induces a lot of noise in the calculation of the transition density due to the low excitation probability for this direction. Figure \ref{fig:td5} and \ref{fig:td6} show the transition densities for the second bright state (around 4.8 eV) obtained from  LR-TDDFT and P-TDDFT, respectively. For the latter, we did a propagation run with the electric field along the y axis, which in this case corresponds to the maximum projection of the transition dipole moment (Table \ref{tbl:casida_excitations}). 

Again, the perfect match between the transition densities computed for the first two bright excitations of the isolated benzaldehyde molecule using different methods, LR-TDDFT and P-TDDFT, validates our new methodology. 

\subsection{Dimers}
Now we can try to extract the information on the exciton dynamics in a more complex system consisting of two molecules applying the \textit{local density analysis} procedure to get the local transition densities. Once the transition densities for each chromophore are obtained, the exciton coupling is computed using eq \ref{eq:tdcm}.

\begin{figure}[h!]
  \centering
   \begin{subfigure}{.475\textwidth}
   \centering
   \includegraphics[width=1.\textwidth]{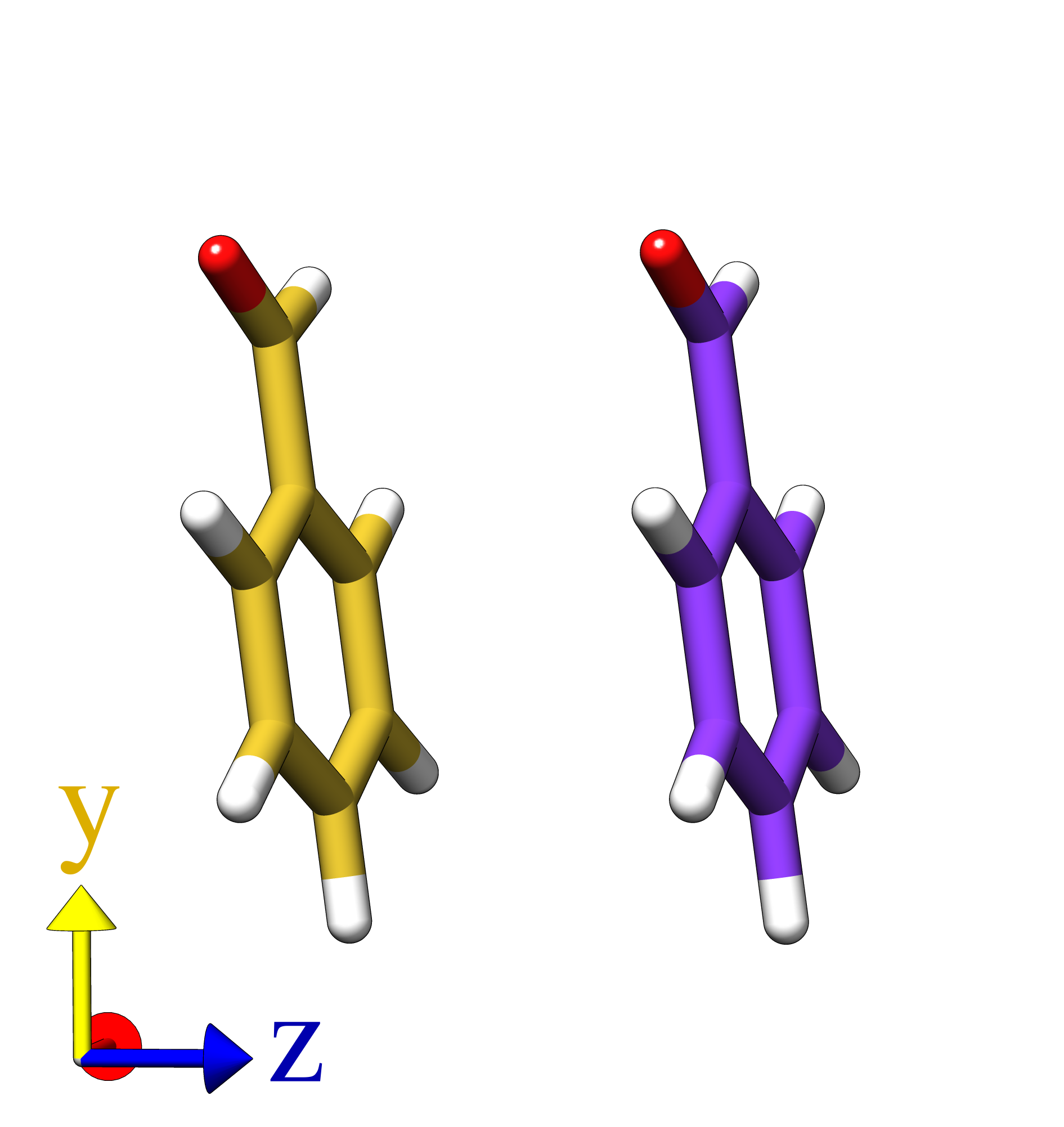}
   \caption{}
   \label{fig:bnzld_dimer}
 \end{subfigure} 
  \begin{subfigure}{.475\textwidth}
  \centering
  \includegraphics[width=1.\textwidth]{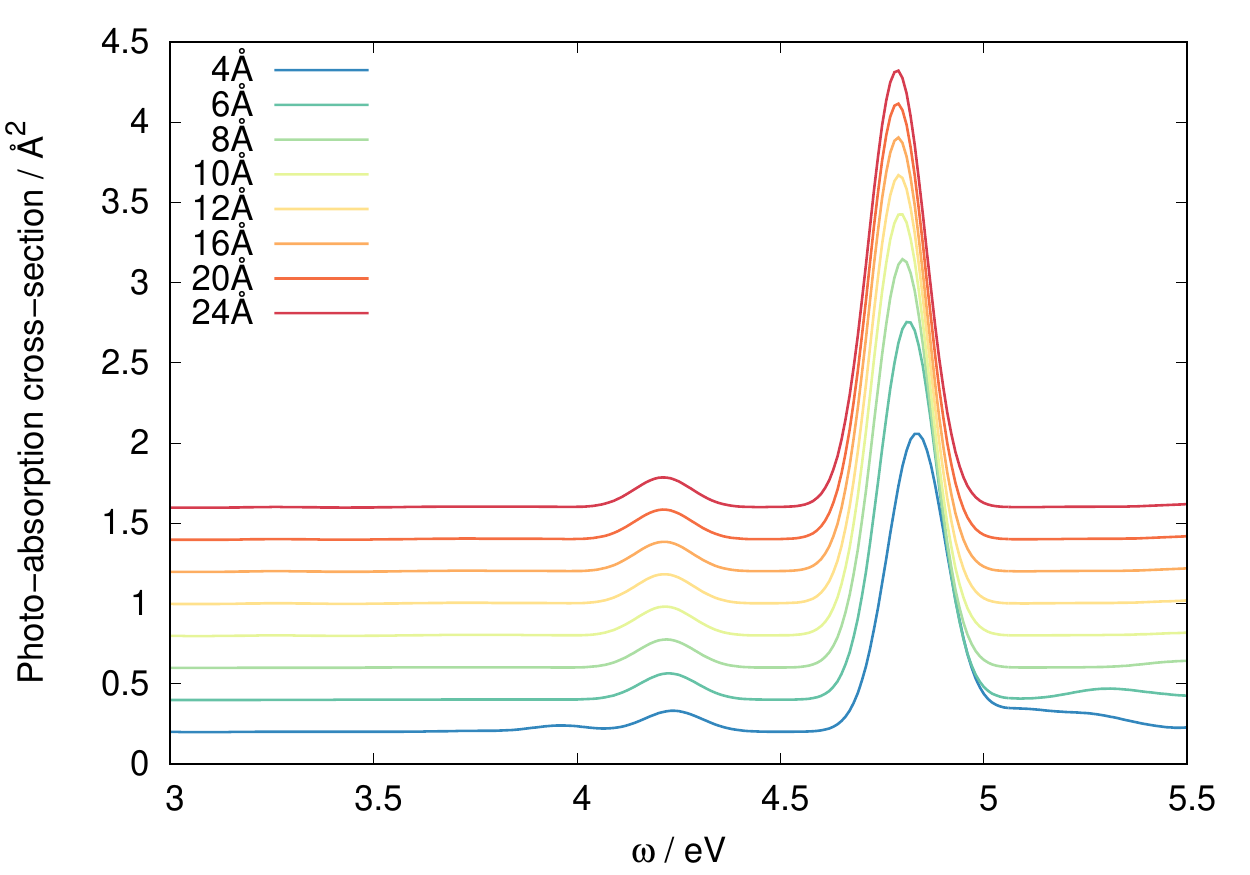}
  \caption{}
  \label{fig:dimer_spectrum}
  \end{subfigure}
  \\
  \begin{subfigure}{.475\textwidth}
  \centering
  \includegraphics[width=1.\textwidth]{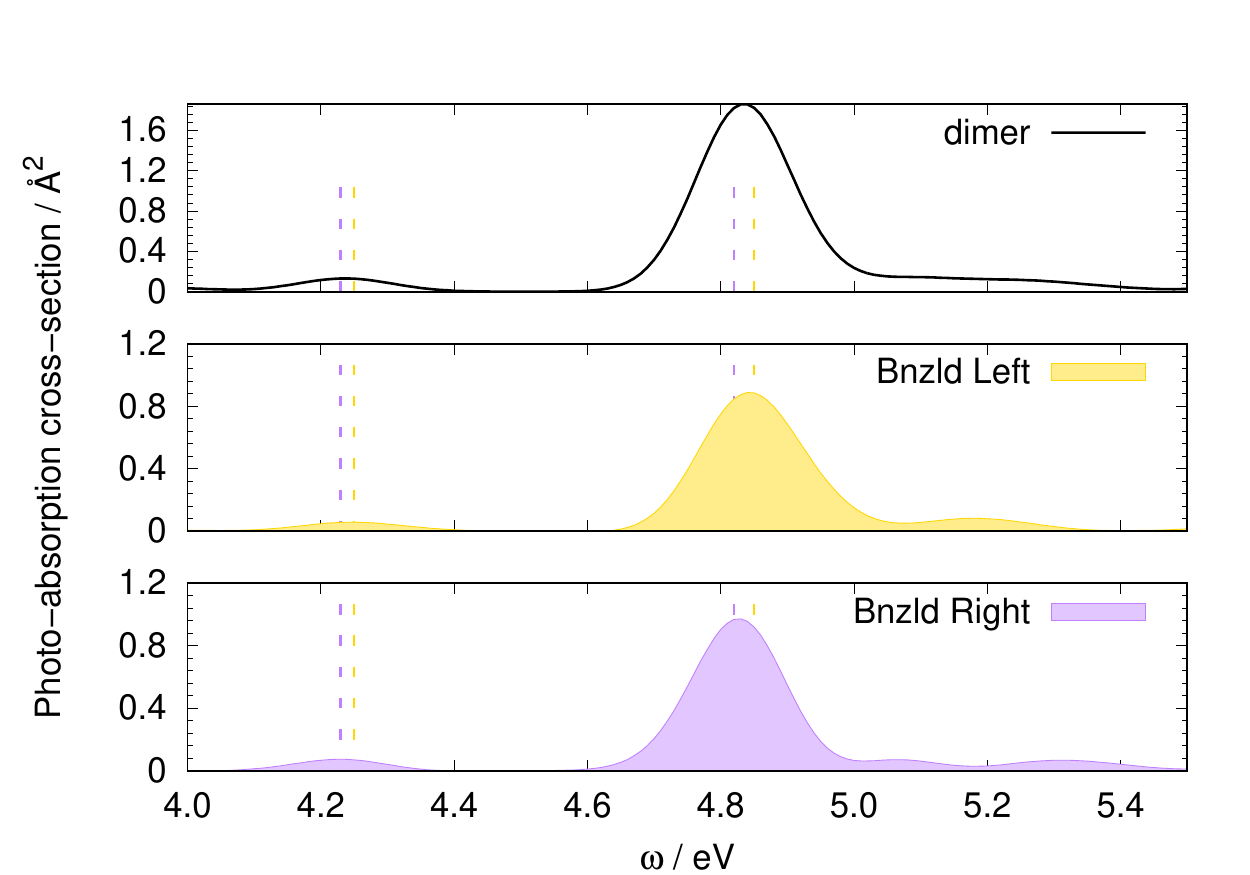}
  \caption{}
  \label{fig:dimer_local_analysis}
  \end{subfigure} 
  \begin{subfigure}{.475\textwidth}
  \centering
  \includegraphics[width=0.7\textwidth, angle=-90]{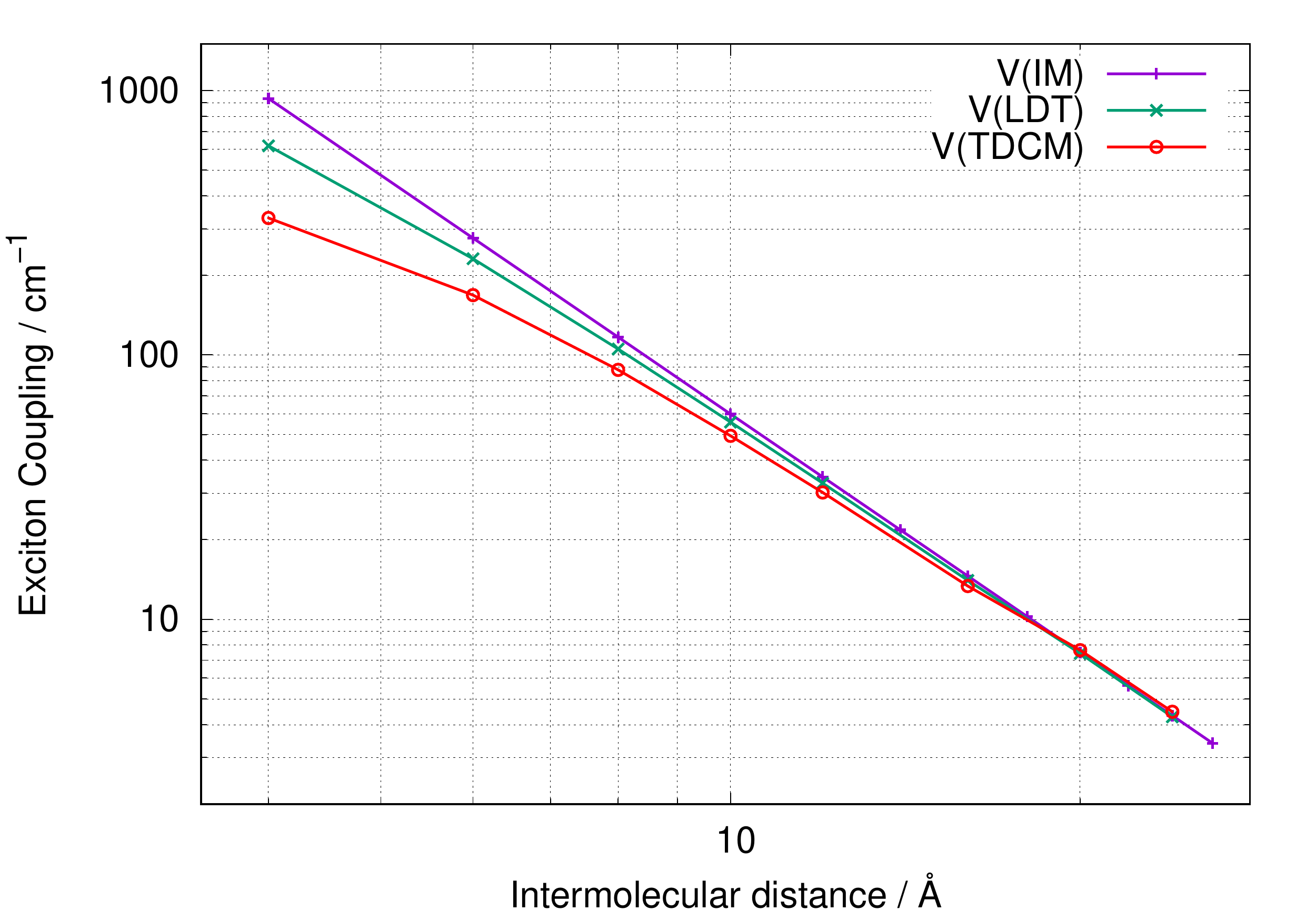}
  \caption{}
  \label{fig:dimer_exciton_coupling} 
 \end{subfigure}
  \caption{(a) Geometry of benzaldehyde dimers. The red, yellow and blue vectors represent the XYZ Cartesian axes respectively. We call "Bnzld Left"  the benzaldehyde molecule in gold with a smaller z coordinate (left), and "Bnzld Right" the purple molecule (right). 
  (b) Comparison of the simulated absorption spectra for benzaldehyde dimers for different separations of the molecules along the z axis. (c) Spectrum decomposition using the \textit{local density analysis} for the dimer with the separation along the z axis of 4 \AA. (d) Exciton coupling evaluated for the strongest excitation around 4.8 eV for  different benozaldehyde dimers.}
  \label{fig:dimer_study}
\end{figure}

Let us exemplify this method by considering the exciton coupling for the face-to-face benzaldehyde dimer as a function of its separation (Figure \ref{fig:bnzld_dimer}). We place two benzalhehyde molecules at distances along the z axis ranging from 4 \AA~to 24 \AA. For each of these dimers, we perform P-TDDFT calculations to get photo-absorption cross-sections (see Figure \ref{fig:dimer_spectrum}).  The effect of the inter-molecular distance is immediately observed in the photo-absorption spectrum. We can see that as the molecules approach each other, a small blue shift develops on the higher peak (around 4.8 eV). There is also a transfer of the oscillator strength at low distances for the low-lying states (around 4.2 eV) and a small peak appears below 4 eV for the very close distance. 

The contributions of each individual molecule to the total spectrum can be obtained performing the \textit{local density analysis}. Figure \ref{fig:dimer_local_analysis} shows the contributions of each molecule for the dimer with the separation along the z axis of 4 \AA.  Due to the small torsion with respect to the xy plane and z axis, similar atoms of the molecules are not perfectly faced in and have a different environment. This results in a slightly different excitation energy for each molecule, shown in the figure by dashed lines. 
After the reconstruction of the individual transition densities with account of the effect of the other molecule, we compute the excitation coupling in the IDA approximation (eq \ref{eq:fret}) and using the TDC method (eq \ref{eq:tdcm}). Figure \ref{fig:dimer_exciton_coupling} shows the variation of the exciton couplings obtained using different methods as functions of the separation along the z axis. From comparison of the couplings V(LDT) and V(TDCM) we see that for benzaldehyde dimers, the IDA validity is limited to separations above 10 \AA, in good agreement with the results shown by Hoffman et al.\cite{Hofmann2010}. This demonstrates the relevance of counting the spatial distribution of the transition density in calculations of the exciton coupling.

The deviation between the exciton couplings computed using the transition dipole moments of the isolated molecules (V(IM)) [$\epsilon$ = 4.78 eV, $\mu(3'A^{1})$ = (-0.0052 \AA, -0.7293 \AA, 0.0940 \AA)] and transition dipole moments computed from P-TDDFT (V(LDT)) (Table \ref{tbl:ldt_distances}) reveals that the electron excitation nature is slightly modified below 8 \AA , i.e. the direction and strength of the transition dipole moment are changed  due to the molecule-molecule interaction.  

\begin{table}[h!]
    \centering
    \begin{tabular}{ c | c | c | c | c | c}
    $R$ &$\epsilon$ & $\mu_{x}$ & $\mu_{y}$ & $\mu_{z}$  & $f_{osc}$ \\
    \hline
4 &   4.85 &-0.0128 &-0.5974 & 0.0722  & 0.1537 \\ 
   &   4.82 &-0.0102 &-0.5994 & 0.0963  & 0.1556 \\
6 &   4.81 & 0.0088 & 0.6651 &-0.0897  & 0.1896 \\
   &   4.81 & 0.0071 & 0.6678 &-0.0902  & 0.1911 \\
8 &   4.80 &-0.0063 &-0.6930 & 0.0924  & 0.2053 \\
   &   4.80 &-0.0059 &-0.6934 & 0.0919  & 0.2055 \\
10 &   4.79 & 0.0051 & 0.7048 &-0.0928  & 0.2118 \\
   &   4.79 &-0.0052 &-0.7050 & 0.0924  & 0.2119 \\
12 &   4.79 &-0.0049 &-0.7105 & 0.0929  & 0.2152 \\
   &   4.79 &-0.0049 &-0.7103 & 0.0927  & 0.2151 \\
16 &   4.79 &-0.0047 &-0.7153 & 0.0930  & 0.2181 \\
   &   4.79 &-0.0047 &-0.7152 & 0.0929  & 0.2180 \\
20 &   4.79 &-0.0047 &-0.7276 & 0.0944  & 0.2256 \\
   &   4.79 &-0.0047 &-0.7277 & 0.0943  & 0.2256 \\
24 &   4.78 &-0.0048 &-0.7284 & 0.0943  & 0.2256 \\
   &   4.78 &-0.0048 &-0.7286 & 0.0944  & 0.2257 \\

    \end{tabular}
    \caption{Comparison of excitation energies ($\epsilon$, in eV), components of the transition dipole moment ($\mu_{x}, \mu_{y}, \mu_{z}$, in \AA) and oscillator strength ($f_{osc}$) for the molecules in the benzaldehyde dimers with different separations $R$ (in \AA) along the z axis.}
    \label{tbl:ldt_distances}
\end{table}

Table \ref{tbl:vcoupling_478ev} shows the values of the computed exciton coupling for the different methods. Columns 6 and 5 show the contributions of the Coulomb and exchange-correlation (XC) terms, respectively, to the coupling V(TDCM) (eq \ref{eq:tdcm}). We see that the F\"orster mechanism coming from the Coulomb interaction governs at the considered distances and it is not until very close distances (below 6 \AA) that the Dexter mechanism related to the exchange-correlation effects becomes non-negligible. This fact is also attributed to the use of the semi-local GGA exchange-correlation functional such as PBE.\cite{Perdew1996,Perdew1997} We expect that the use of long-range XC functionals\cite{Bruckner2017,Stehr2014, Hirata2001, Stehr2016} modifies this trend and the XC term should become more significant also at larger distances. 

\begin{table}[h!]
  \centering
  \begin{tabular}{ c | c | c | c | c | c }
  $R$ &  V(IM)          &     V(LDT)    &     V(TDCM)    &  V\textsubscript{h}          &    V\textsubscript{XC}     \\
   \hline      
   4.0 &933.313  &    618.470 & 329.559  &  328.114 &  1.445 \\
   6.0 &276.537  &    231.079 & 168.345  &  168.3662 &  0.0210 \\
   8.0 &116.664  &    105.256 &  87.874  &   87.874 &  0.000 \\
  10.0 & 59.732  &     55.733 &  49.392  &  -49.392 &  0.000 \\
  12.0 & 34.567  &     32.746 &  30.170  &   30.170 &  0.000 \\
  16.0 & 14.583  &     14.034 &  13.353  &   13.353 &  0.000 \\
  20.0 &  7.467  &      7.420 &   7.640  &    7.640 &  0.000 \\
  24.0 &  4.321  &      4.2745 &   4.470  &    4.470 &  0.000 \\          

  \end{tabular}
  \caption{Exciton couplings V(IM), V(LDT), V(TDCM) (in cm\textsuperscript{-1}) and contributions of the Coulomb (V\textsubscript{h}) and exchange-correlation (V\textsubscript{XC}) terms to V(TDCM) for dimers of benzaldehyde with different separations $R$ (in \AA) along the z axis. }
  \label{tbl:vcoupling_478ev}
\end{table}

\subsection{Complex multichromophore system}

\begin{figure}[h!]
\centering
   \begin{subfigure}{0.41\textwidth}
      \centering
      \includegraphics[width=0.95\textwidth]{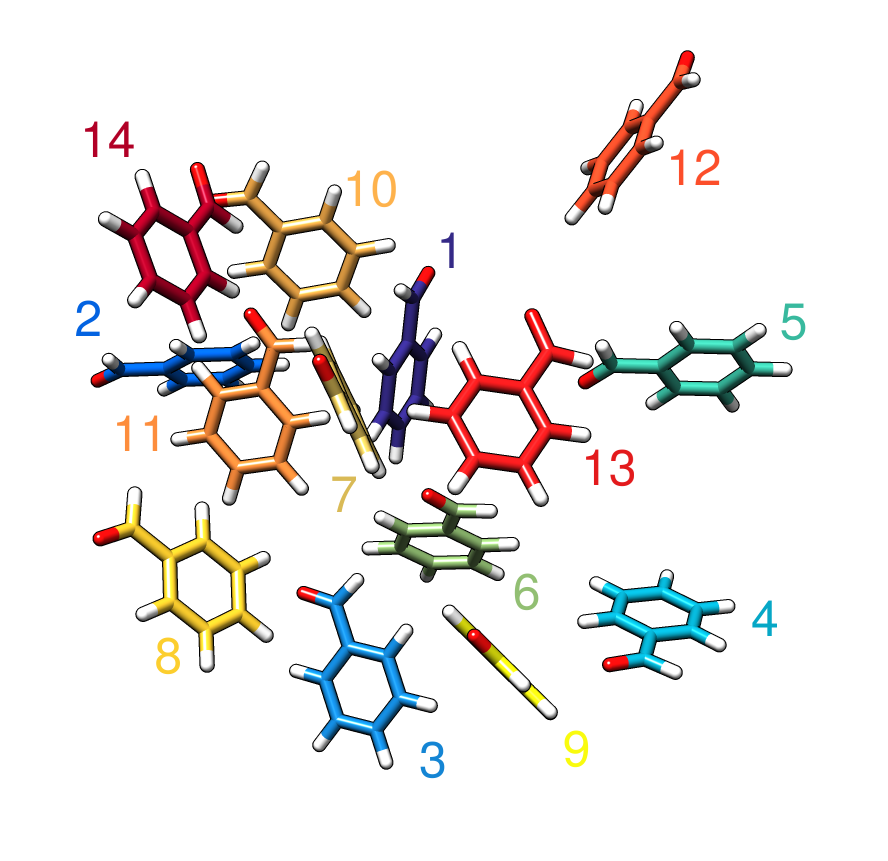}
      \caption{}
      \label{fig:cluster_geometry}
   \end{subfigure}
   \begin{subfigure}{0.58\textwidth}
      \centering
      \includegraphics[width=0.95\textwidth]{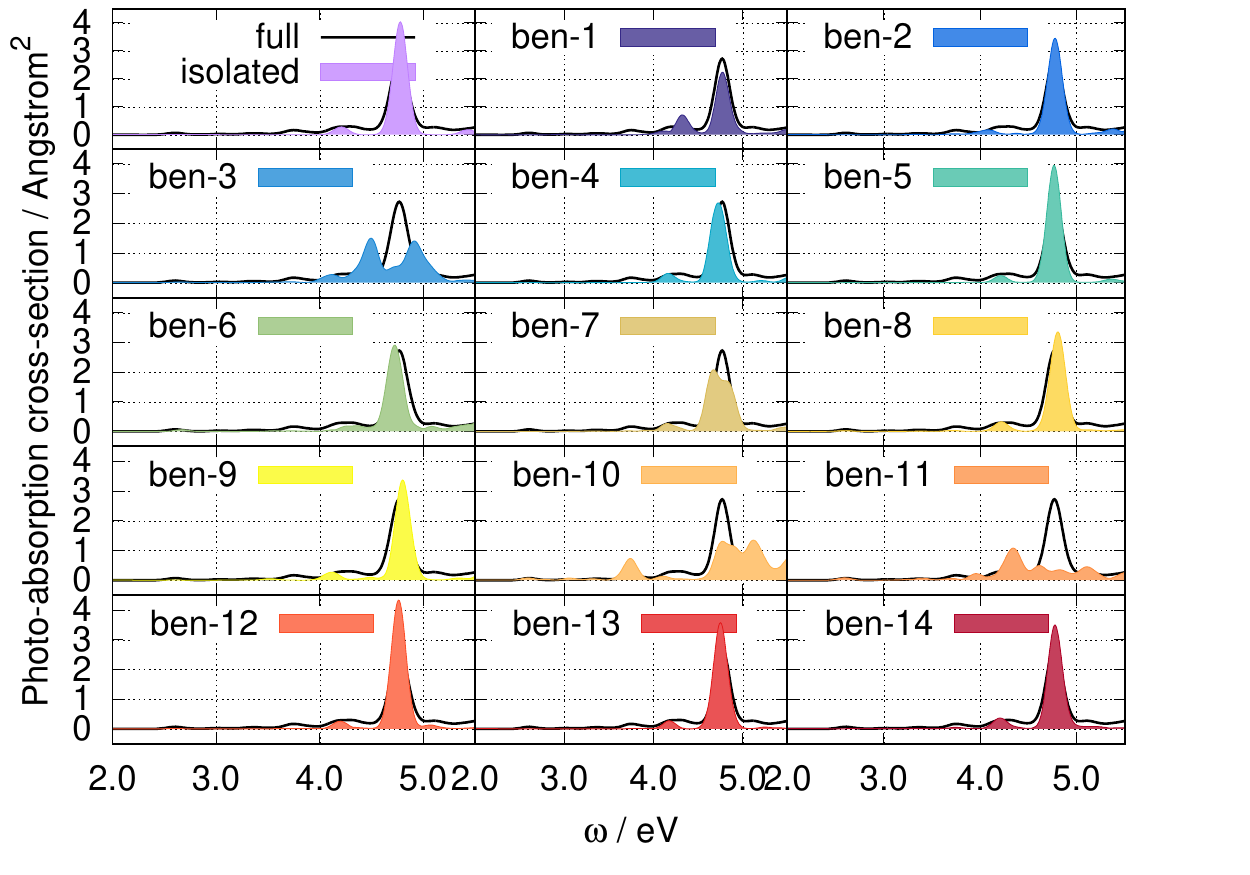}
      \caption{}
      \label{fig:cluster_spectra}
   \end{subfigure}
   \caption{(\ref{fig:cluster_geometry}) Geometrical distribution of  14 benzaldehyde molecules in the cluster. (\ref{fig:cluster_spectra}) Spectrum decomposition for the 14 benzaldehyde cluster.}
   \label{fig:cluster_description}
\end{figure}

Lets now consider a complex system with more than two photo-active molecules. We have created a cluster consisting of 14 randomly oriented benzaldehyde molecules. To ensure a plausible disposition, during the cluster generation, we avoid inter-atomic distances between atoms of different molecules smaller than the sum of the covalent radii plus 30\% of the corresponding radii. Figure \ref{fig:cluster_geometry} shows the spacial distribution of the 14 molecules in the cluster with the distances between centers-of-mass of the molecules ranging between 5.17 \AA~and 19.86 \AA.

After the calculation of the ground state for the full system, the \textit{local density analysis} is carried out in order to determine the charge density belonging to each molecule. The appropriateness of each local density domain is validated by its total charge. It is checked that the valence charge corresponds to 40 electrons for each molecule, with the maximum deviation found of 0.003 electron charge (which corresponds to the error of 0.00075\%). Then we perform three P-TDDFT calculations with the electric field perturbation applied along each of the Cartesian axes. We propagate the system for the total propagation time of 40 fs. For each local domain, the  dynamic polarizability is calculated from the local induced dipole moment, and therefore the local contribution of each molecule to the global absorption spectrum is obtained (see Figure \ref{fig:cluster_spectra}). Besides, as described above, the transition dipole moments for a each peak are obtained by fitting the diagonalized dynamic polarizability tensor  (see Table 
\sref{SItbl:cluster_mu_fit}). Finally, we compute the transition density for a given excitation for each molecule of the cluster defined for the corresponding charge density domain using eq \ref{eq:n0j-PTDFT}. 
The computed transition densities at the most intense peak below 5 eV are influenced by the presence of the embedding environment (see Figure \sref{SIfig:cluster_tp}). The effects due to spurious contamination of near located bright states can be evaluated by comparing the transition dipole moments obtained by the Gaussian fit and by applying the dipole operator over the transition density, analogously to the eq \ref{eq:tm_from_tp}. Tables \sref{SItbl:cluster_mu_ida}-\sref{SItbl:cluster_mu_ftp} show this comparison, which corresponds to a mean deviation of 0.02 \AA~in the transition dipole moments, and a mean angle twist less than 6\degree .

\begin{figure}[!h]
  \centering
  \begin{subfigure}{0.45\textwidth}
    \centering
    \includegraphics[width=0.95\textwidth]{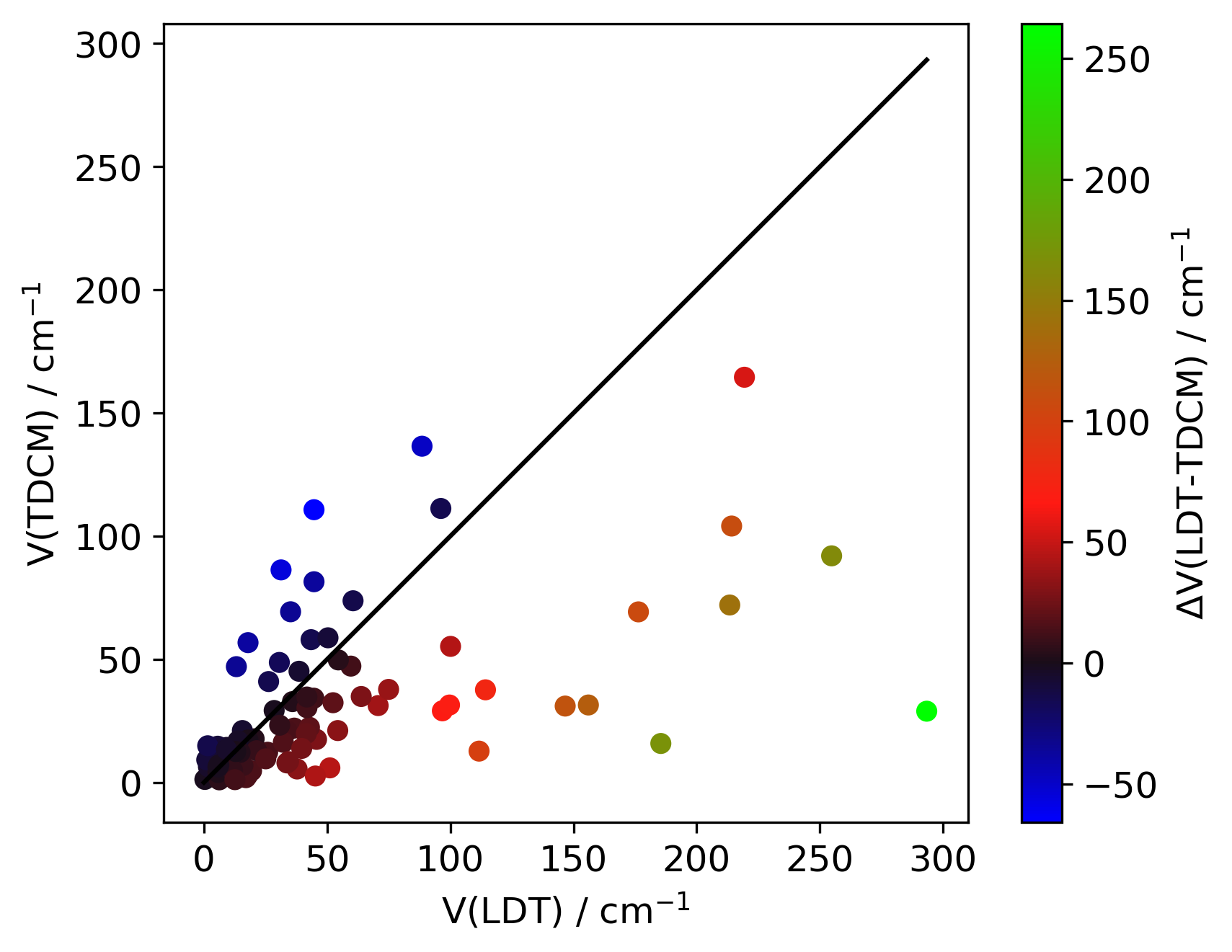}
    \caption{}
    \label{fig:cluster_vvalues}
  \end{subfigure}
  \begin{subfigure}{0.45\textwidth}
    \centering
    \includegraphics[width=0.95\textwidth]{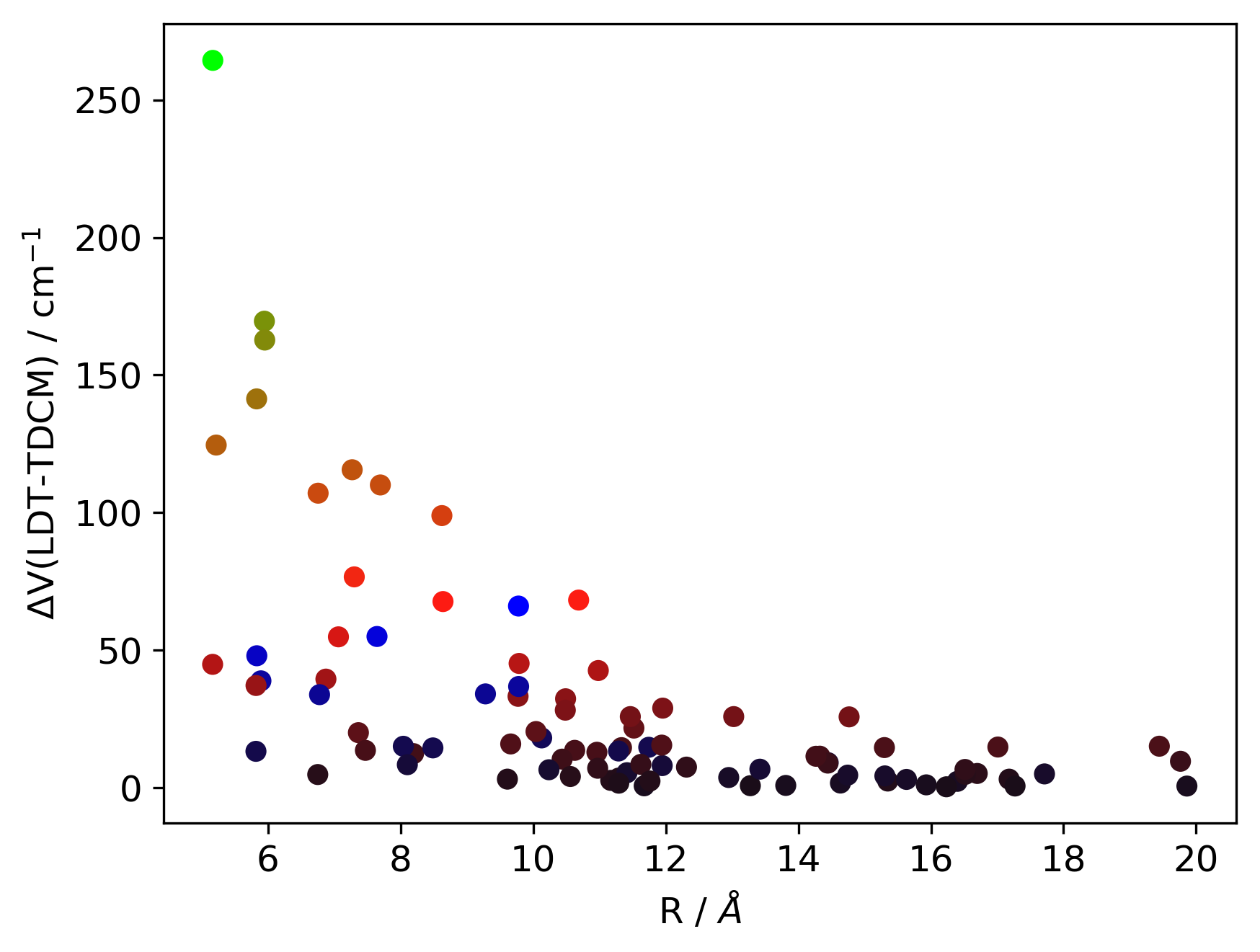}
    \caption{}
    \label{fig:cluster_r_vdiff}
  \end{subfigure}
  \caption{(\ref{fig:cluster_vvalues}) Correlation between the values computed using the transition density cube method (V(TDCM)) versus the values computed from the ideal dipole approximation using the transition dipole moments found for local domains from the fit of the diagonalized dynamic polarizability tensor (V(LDT)). (\ref{fig:cluster_r_vdiff}) Difference between V(LDT) and V(TDCM) as a function of the center-of-mass distance.}
 
  \label{fig:cluster_couplings}
\end{figure}

We compute now the exciton coupling for the entire system using the different approaches described above. A comparison of the values obtained for all pairs exciton coupling are specified in Tables \sref{SItbl:exc_coupling_lfret}-\sref{SItbl:exc_coupling_ida}.
Figure \ref{fig:cluster_couplings} shows the differences in the exciton couplings computed using TDCM (eq \ref{eq:tdcm}) and LDT (eq \ref{eq:fret}), i.e. the transition densities from eq \ref{eq:n0j-PTDFT} and the transition dipole moments from the fit and diagonalization of the dynamic polarizability tensor (eq \ref{eq:spectralfunction}).
The difference between both methods is not homogeneous and strongly depends on the specific pair of molecules.

We can see from Figure \ref{fig:cluster_r_vdiff} that the ideal dipole approximation is valid for the intermolecular distances (distances between the centers of mass) larger than 12 \AA. Hence the excitation energy transfer between those molecules can be explained by the F\"orster resonant mechanism. 
However, at lower distances, this approximation fails 
as the V(TDCM) and V(LDT) values largely differ.  This fact demonstrates the need for the method that enables accurate calculations of transition densities for individual subunits by treating all the system at the same level of theory.

\section{CONCLUSIONS}

In this work, we demonstrate that treating all the complex photoactive system at the \textit{ab initio} level of theory allows accurate calculations of the exciton dynamic properties. Although our approach suffers from intrinsic limitations of some exchange-correlation functionals in the TDDFT method such as the proper description of charge-transfer states, it can provide a more accurate alternative to other approaches for weakly interacting large systems by including explicitly the quantum mechanical environment effects.

In this paper we derived an analytic method to obtain transition dipole moments and transition densities from time-propagation TDDFT calculation in the linear-response regime (P-TDDFT).
We validated that the transition dipole moment can be obtained by Gaussian fitting of the absorption spectra and its direction is recovered from the eigenvectors of the dynamic polarizability diagonalization. 
We proved that the transition density for a given excitation can be accurately recovered from the Fourier transform of the time-dependent response density using the previous knowledge of the corresponding transition moment. 

More interesting, we introduced the \textit{local density analysis} to determine the transition properties for multichromophore systems. This procedure allows to accurately extract transition dipole moments and transition densities for individual molecules taking into account all effects due to their environment, since the entire system is modeled at the same level of theory. The calculation of these properties enables studies of exciton coupling in complex systems, a key parameter to understand the energy transfer processes taking place, for example, in photosynthetic light-harvesting complexes or the light-emitting layer of OLED devices. 

This method also makes possible to disentangle the mechanisms of exciton transfer. The contribution of different interactions can easily be evaluated. In this work, we included the Coulomb and Exchange-Correlation effect (for density-dependent XC functionals), but other mechanisms such as an explicit terms due to the polarizable environment\cite{Hsu2001,Curutchet2009} can be considered as well. 

It is important to mention that this method is implemented into the open-source \texttt{OCTOPUS} package, which provides a platform for performing P-TDDFT in the real-space basis. It is highly parallelized and is therefore suitable for efficient calculations of exciton couplings in large and complex systems.

\begin{acknowledgement}

JJS and IL thank the European Research Council (ERC-2010-AdG-267374), Spanish grant (FIS2016-79464-P), and Grupos Consolidados (IT578-13) and EU-H2020 project ``MOSTOPHOS'' (n. 646259) for the financial support. JJS gratefully acknowledges the Spanish grant IJCI-2014-22204, and H2020-EINFRA-5-2015 project ``NOMAD'' (n. 676580) and the funding from the European Union Horizon 2020 research and innovation program under the Marie Sklodowska-Curie Grant Agreement No. 795246-StrongLights. The authors gratefully thank Prof. Angel Rubio for his comments and support. 

\end{acknowledgement}

\begin{suppinfo}

\begin{itemize}
  \item{Supplementary Information document includes: Section \sref{SIsec:ComputDetails}. Computational Details: Extended description the computational details used to obtain the \textit{ab initio} results. Section \sref{SIsec:broadening}. Damping factor effect on the peak broadening: analytical derivation on the broadening of the absorption spectrum peaks due to the introduction of a Gaussian damping function on the short-time Fourier transform. Section \sref{SIsec:SupFigs} and Section \sref{SIsec:SupTabs}, supplementary figures and tables that support main text results.}
  
\end{itemize}

\end{suppinfo}


\bibliography{jjs_il_ptddft}

\end{document}